\documentclass[lettersize,journal]{IEEEtran}
\usepackage{url}
\usepackage{hyperref}
\hypersetup{
    colorlinks=true,
    linkcolor=[RGB]{0,70,140},
    urlcolor=[RGB]{0,70,140},
    citecolor=[RGB]{0,70,140}
}
\usepackage{xcolor}
\usepackage{amsmath,amsfonts}
\usepackage{algorithm}
\usepackage{array}
\usepackage[caption=false,font=normalsize,labelfont=sf,textfont=sf]{subfig}
\usepackage{textcomp}
\usepackage{stfloats}
\usepackage{url}
\usepackage{verbatim}
\usepackage{graphicx}
\usepackage{cite}
\usepackage[noend]{algpseudocode}
\usepackage{amssymb}
\usepackage{bm}
\usepackage{makecell}
\usepackage{booktabs,multirow}

\newcommand{\sd}[1]{\textsubscript{$\pm$#1}}
\begin{document}

\title{HiTMS: A High-Throughput Multi-Stream \\ Linguistic Steganography Framework}

\author{Ruiyi Yan, Zhongliang Yang, and Yugo Murawaki
\thanks{This work was supported by JSPS KAKENHI Grant Number JP26KJ1382.}
\thanks{Ruiyi Yan and Yugo Murawaki are with the Graduate School of Informatics, Kyoto University, Kyoto 606-8501, Japan. 
Zhongliang Yang is with School of Cyberspace Security, Beijing University of Posts and Telecommunications, Beijing 100876, China. (E-mail: ruiyi@nlp.ist.i.kyoto-u.ac.jp; murawaki@i.kyoto-u.ac.jp; yangzl@bupt.edu.cn.)}
\thanks{Corresponding author: Yugo Murawaki.}
\thanks{GitHub repository for this work is \url{https://github.com/ryehr/HiTMS_steganography}.}
}

\markboth{UNDER REVIEW, PREPRINT, JULY 2026}{}

\maketitle

\begin{abstract}
Generative linguistic steganography conceals secret bits within the sampling randomness of large language models.
Existing schemes are single-stream, conveying an entire secret through a single response to a single prompt. This convention incurs limitations:  it provides no protocol-level support for batched multi-stream inference, and naive co-batching does not conceal slot occupancy or payload completion.
We propose the High-Throughput Multi-Stream (HiTMS) framework, which distributes a secret across multiple responses produced jointly over successive rounds of interaction. Each round embeds and extracts several streams within a single batched call, thereby amortizing the cost of model invocation and substantially improving throughput. To ensure recoverability, HiTMS wraps each response in a self-describing frame and employs a key-derived schedule that binds streams to slots and fills unused slots with decoys, guaranteeing exact recovery while concealing the number of active streams.
The framework is agnostic to both the language model and the steganographic coder. 
Across eight dataset--model--coder settings, eight-stream HiTMS achieves up to \(4.3\times\) higher embedding and extraction speeds than single-stream baselines, while reducing the average area under the receiver operating characteristic curve (AUROC) of steganalyzers from \(0.681\) to \(0.601\). Experiments with \(4\) to \(64\) streams demonstrate sustained throughput gains as concurrency increases.
\end{abstract}

\begin{IEEEkeywords}
Linguistic steganography, natural language processing, multiple streams, high throughput, imperceptibility.
\end{IEEEkeywords}


\section{Introduction}

Linguistic steganography conceals a secret message within seemingly innocuous natural-language text, so that both the existence of the communication and its content are hidden from observers. Driven by modern large language models (LLMs), linguistic steganography has shifted from editing cover texts~\cite{ueoka-etal-2021-frustratingly, zheng2022autoregressive} to directly steering the token-level sampling process of an autoregressive LM, embedding secret bits into the randomness of generation~\cite{8470163, 9193914,  bai2025shimmer,wang2025sparsampefficientprovablysecure,yan2025comprehensive}. 


Existing schemes define a single payload-bearing stream per protocol instance: one secret is conveyed through one response to one prompt~\cite{10888944,yan-etal-2026-anchored}. This exposes the limitation of forcing linguistic steganography to rely on non-batched inference. Specifically, protocol-level support for batched, multi-stream, multi-round transmission is absent. A serving system could co-batch multiple independent single-stream instances as an implementation-level optimization; however, such naive co-batching does not define how dynamically completing streams should be framed, scheduled, and recovered across rounds. It may also expose slot occupancy and stream completion through changing batch sizes or response lengths. These limitations motivate a protocol that supports batched inference while preserving recoverability and concealing the internal stream pattern.

Motivated by these limitations, we propose \textbf{HiTMS}, a \textbf{Hi}gh-\textbf{T}hroughput \textbf{M}ulti-\textbf{S}tream linguistic steganography framework that fragments and conveys secret messages across LM batch-inference responses over multiple rounds of interaction. 
\textbf{To the best of our knowledge, we are the first to formally propose multi-stream linguistic steganography based on autoregressive LMs.} It is a significant step from isolated single-shot transmission toward a \emph{linguistic version of steganography as a service (LSaaS)}~\cite{5166821, 8646434, 9004347, 11524553}, because it supports concurrency for multiple requests.

\begin{figure}[t]
\centering
\includegraphics[width=\linewidth]{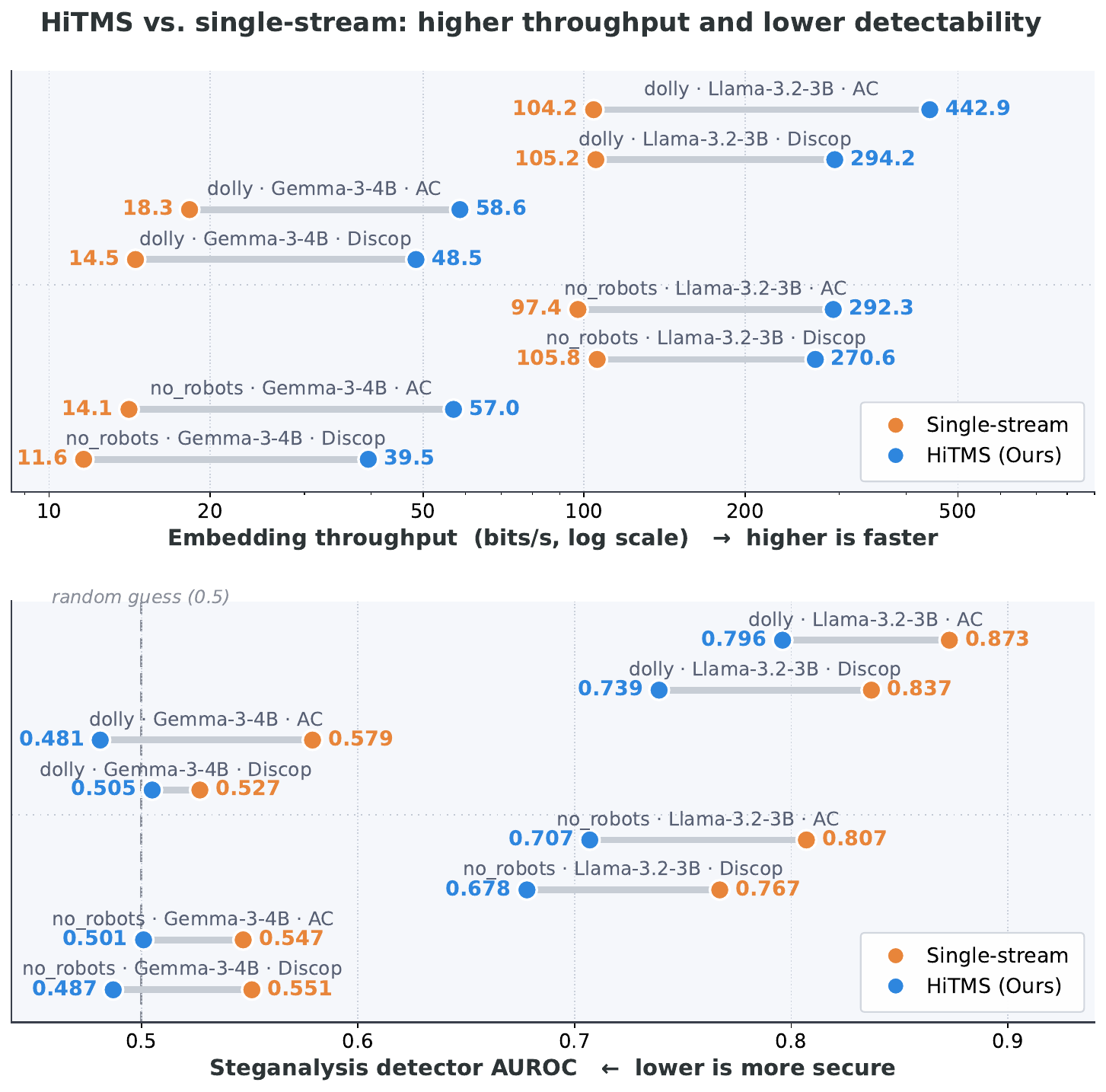}
\caption{Comparison of eight-stream HiTMS with the basic single-stream baselines across eight dataset--model--coder settings. Top: embedding throughput in bits/s on a logarithmic scale, where higher is better. Bottom: AUROC of the BERT-base steganalyzer, where lower is more secure and \(0.5\) denotes random guessing. Each line connects the two methods under the same setting; HiTMS consistently achieves higher throughput and anti-steganalysis capability.}
\label{fig:teaser}
\end{figure}

\begin{figure*}[t]
    \centering
        \includegraphics[width=0.9\textwidth]{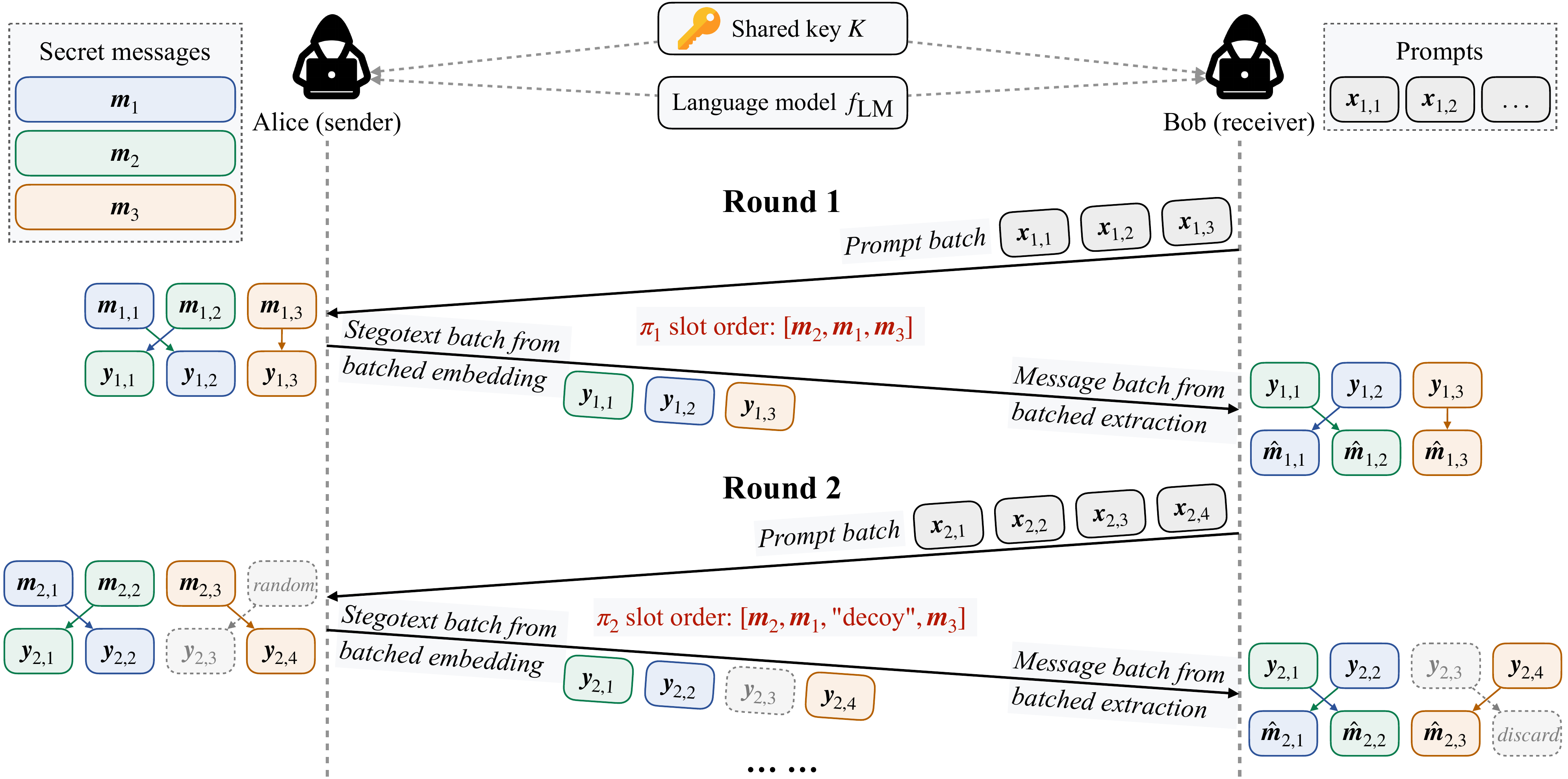}
            \caption{Overview of HiTMS. In each round, Bob sends a batch of prompts and Alice returns a stegotext (response) batch produced by one batched LM call. A key-derived mapping assigns active secret streams to response slots, while unused slots carry decoys. The same-colored fragments indicate that one secret stream can span multiple rounds until Bob recovers all payloads.}
    \label{fig:main}
\end{figure*}

Our contributions are summarized as follows. First, we formulate multi-stream, multi-round linguistic steganography and propose HiTMS, a model- and coder-agnostic framework that fragments secret messages across multiple response streams and processes each round using a single batched LM call, thereby amortizing inference costs and enabling concurrent steganographic transmission.
Figure~\ref{fig:teaser} summarizes its main advantages: compared with the single-stream baselines, eight-stream HiTMS  achieves substantially higher throughput (with speedups of up to \(4.3\times\)) and lower steganalyzer AUROC (from \(0.681\) to \(0.601\) on average) across all eight dataset--model--coder settings and various steganalyzers.
Besides, as illustrated in Figure~\ref{fig:main}, we design a self-describing framing and scheduling protocol that combines encrypted headers, key-derived stream-to-slot mapping, filler bits, and decoy responses. These mechanisms ensure exact recovery across variable-length responses and multiple rounds while concealing the number and arrangement of active streams. We conduct extensive experiments across two datasets, two LMs, and two steganographic coders.
The scalability experiments with \(4\) to \(64\) streams under both arithmetic coding (AC)~\cite{ziegler-etal-2019-neural} and Discop~\cite{10179287} confirm that its throughput gains persist as concurrency increases.

\section{Preliminaries and Notations}
\label{sec:formulation}

\subsection{Language Model Preliminaries}

An autoregressive language model (LM) operates over a finite vocabulary
$\mathcal{V}$. 
Consider a sequence of $T$ LM-generated tokens
$[s^{(0)},\ldots,s^{(T-1)}]\in\mathcal{V}^{T}$. The generated sequence is
preceded by a \emph{prompt} (commonly an instruction or a question)
$[s^{(-N_p)},\ldots,s^{(-1)}]$ of length $N_p$, for which we use negative
indices to distinguish the prompt tokens from the generated tokens.
Generation proceeds autoregressively. At each position $t$, the LM takes the
prompt together with all previously generated tokens as its context and
outputs the next-token distribution
\begin{equation}
    \bm{p}^{(t)}
    \;=\;
    \bigl(p^{(t)}_{1},\ldots,p^{(t)}_{|\mathcal{V}|}\bigr)
    \;=\;
    f_\text{LM}\bigl(\cdot\mid
    s^{(-N_p)},\ldots,s^{(t-1)}\bigr),
\end{equation}
where $p^{(t)}_v$ is the probability assigned to the $v$-th vocabulary token.
The vector $\bm{p}^{(t)}$ is obtained by applying a softmax to the logits
produced by the LM head and therefore satisfies
$p^{(t)}_v\geq 0$ and $\sum_{v=1}^{|\mathcal{V}|}p^{(t)}_v=1$. A decoding
strategy then selects $s^{(t)}$ from this distribution. Standard covertext
generation can use multinomial sampling, temperature scaling, top-$k$ or
top-$p$ filtering. 
The selected token is appended to the context before the distribution
for position $t+1$ is evaluated. Consequently, the distribution at every
position depends on the token prefix generated up to that point.

This token-by-token paradigm is also the basis of LM-based linguistic
steganography. Instead of using unconstrained random bits to sample from
$\bm{p}^{(t)}$, a steganographic coder uses secret bits to resolve the sampling
choice. 
In the setup of \textit{symmetric} linguistic steganography introduced below, the coder must expose the same token to the LM on both
the sender and receiver sides, so that both parties can reconstruct the same
sequence of conditional distributions.

\subsection{Single-Stream LM-Based Linguistic Steganography}

Single-stream generation is the de facto standard in generative linguistic
steganography over autoregressive LMs~\cite{ziegler-etal-2019-neural,
10179287, wang2025sparsampefficientprovablysecure,
yan-murawaki-2026-efficient}, apart from schemes built on diffusion
LMs~\cite{10.1007/978-981-96-7005-5_12, qi2026stead}. In this mode, one secret
bitstring is conveyed through the token-level sampling randomness of one LM
response.

Let $\mathcal{B}=\{0,1\}$. For a bitstring $\boldsymbol{a}$,
$\boldsymbol{a}[u{:}v]$ denotes its $u$-th through $(v{-}1)$-th bits under
0-based indexing, while $\boldsymbol{a}[u{:}]$ denotes the suffix beginning at
position $u$. Concatenation of two bitstrings is denoted by $\Vert$. 
Alice, the sender, and Bob, the receiver, share a master key \(K \in \{0,1\}^{\kappa}\) established through an offline secure channel, in which \(\kappa\) denotes the key length (security parameter), and \(K\) is a \(\kappa\)-bit binary string.
They also use the same LM, tokenizer, prompt, and
steganographic coder, which allows Bob to replay the generation process from
the observed token sequence.

Given a secret $\boldsymbol{m}\in\mathcal{B}^{L}$ and a prompt, Alice first
computes the LM distribution for the current position and then invokes a
token-level embedder:
\begin{equation}
\label{eq: token-level embedder}
    s^{(t)}
    \;=\;
    \mathsf{Emb}\bigl(\boldsymbol{p}^{(t)},\,
    \boldsymbol{m},\,K,\,t\bigr),
    \qquad t=0,1,\ldots,T-1,
\end{equation}
where
$\boldsymbol{p}^{(t)}=f_\text{LM}(\cdot\mid
s^{(-N_p)},\ldots,s^{(t-1)})$. The interface $\mathsf{Emb}$ is
\emph{stateful}: it maintains a pointer to the first unconsumed position of
$\boldsymbol{m}$ and updates that pointer after every token. Depending on the
shape of $\boldsymbol{p}^{(t)}$ and on the chosen coding rule, a single token
may consume several bits, one bit, or no bit. 
Bob receives the prompt and the complete stegotext response. At position $t$,
he conditions the same LM on the prompt and the already observed stegotext
prefix, thereby reconstructing the same $\boldsymbol{p}^{(t)}$ used by Alice.
He then applies the matching token-level extractor,
\begin{equation}
\label{eq: token-level extractor}
    \hat{\boldsymbol{m}}^{(t)}
    \;=\;
    \mathsf{Ext}\bigl(\boldsymbol{p}^{(t)},\,
    s^{(t)},\,K,\,t\bigr)
    \;\in\;\mathcal{B}^{*},
    \qquad t=0,1,\ldots,T-1,
\end{equation}
where $\hat{\boldsymbol{m}}^{(t)}$ is the possibly empty substring represented
by token $s^{(t)}$ under the coder state at position $t$. The recovered message
is obtained in token order as
$\hat{\boldsymbol{m}}=
\hat{\boldsymbol{m}}^{(0)}\Vert\cdots\Vert
\hat{\boldsymbol{m}}^{(T-1)}$. The scheme is \emph{lossless} if
$\hat{\boldsymbol{m}}=\boldsymbol{m}$ whenever Alice and Bob begin with the
same inputs and synchronized coder states.

We treat $\mathsf{Emb}$ and $\mathsf{Ext}$ as abstract steganographic coder
interfaces. Concrete instantiations include arithmetic coding
(AC)~\cite{ziegler-etal-2019-neural}, adaptive dynamic grouping
(ADG)~\cite{zhang-etal-2021-provably}, distribution-preserving
Discop~\cite{10179287}, and others. These coders differ in how they partition
or traverse $\boldsymbol{p}^{(t)}$, and therefore offer different trade-offs
among embedding capacity, computational cost, and distributional fidelity.
The framework designed below does not rely on the internal operation of any
one coder; it only requires a matched, lossless pair of interfaces.


\section{HiTMS: High-Throughput Multi-Stream Linguistic Steganography}

Single-stream steganography assumes that an entire secret is conveyed through
one LM response. When several secrets must be delivered, this convention
invokes the LM on one prompt at a time and serializes generated responses. It therefore underutilizes the batched-inference capability and limits the number of secret bits that can be
processed within a fixed time budget. Moreover, a rigid one-question--one-answer
exchange does not reflect realistic LM usage, which can involve multiple
concurrent and successive interactions~\cite{cheng-etal-2023-batch}.

To address these limitations, we propose \textbf{HiTMS}, a high-throughput
multi-stream linguistic steganography framework. HiTMS fragments
secret messages across multiple LM responses and, when necessary, across
multiple rounds of interaction. In each round, several response streams are
generated together in one batched LM call. A key-derived schedule determines
which secret stream occupies each response slot, while encrypted headers,
filler bits, and decoy responses allow Bob to recover every fragment without
revealing the number or arrangement of active streams. HiTMS changes how
secret bits are framed and scheduled around an LM-based coder; it does not
change the token-level coding rule itself.

\subsection{Design Objectives}
\label{sec:objectives}

Generalizing single-stream steganography to a multi-stream, multi-round setting
introduces challenges that do not arise when one response carries one complete
message. A stream may span several responses of different realized lengths,
the set of unfinished streams may change after every round, and a batch may
contain more slots than currently active streams. HiTMS is designed around the
following four objectives.

\paragraph{Compatibility with batched inference}
All responses within a round should be produced by a single batched LM call.
At each autoregressive step, the LM evaluates the current prefixes of multiple
slots together, allowing the dominant inference cost to be shared across the
batch. The framing and scheduling operations should add only lightweight work
around the underlying coder. As a result, within a given time budget, HiTMS
should convey substantially more secret bits than repeatedly invoking a
single-stream scheme.
For batched LM inference, let \(\boldsymbol{X}=\{\boldsymbol{x}_j\}_{j=1}^{n}\) be a batch of prompts, where
\(\boldsymbol{x}_j=[s_j^{(-N_{p,j})},\ldots,s_j^{(-1)}]\).
The generated prefix in slot \(j\) is \([s_j^{(0)},\ldots,s_j^{(t-1)}]\), and let \(\mathcal{J}_t\subseteq\{1,\ldots,n\}\) denote the slots that have not terminated before decoding position \(t\). A single batched forward pass computes
\begin{equation}
\begin{aligned}
 \left(\boldsymbol{p}_j^{(t)}\right)_{j\in\mathcal{J}_t}
 &=
 f_{\mathrm{LM}}^{\mathrm{batch}}\!\left(
   \left(s^{(-N_p,j)}_j,\ldots,s^{(t-1)}_j\right)_{j\in\mathcal{J}_t}
 \right) .
\end{aligned}
\label{eq:batched-lm-inference}
\end{equation}
Here, \(f_{\mathrm{LM}}^{\mathrm{batch}}\) denotes a batched evaluation of the same LM \(f_{\mathrm{LM}}\). A decoding strategy then selects \(s_j^{(t)}\) from \(\boldsymbol{p}_j^{(t)}\) for every \(j\in\mathcal{J}_t\) in parallel. Padding and attention masks, omitted from Equation~\eqref{eq:batched-lm-inference}, align prefixes of different lengths. 

\paragraph{Recoverability of each stream in each round}
Every fragment must be assigned to the correct stream and placed at the correct
offset within that stream. This is nontrivial for two reasons. First, the
number of secret bits carried by a response depends on the response's realized
token sequence and length, so Bob cannot know the fragment size in advance.
Second, the active subset of streams is dynamic: a stream leaves the active set
as soon as all of its bits have been delivered, whereas a stream that is not
served in the current round must remain pending. The protocol must therefore
be \emph{self-describing}. From the shared state, key, and information carried
by each served stegotext, Bob must be able to reproduce the stream-to-slot
assignment, determine which extracted bits are payload, update the correct
stream offset, and detect when all streams have completed.

\paragraph{Indistinguishability of stream patterns}
The number of prompts $n_r$ in a round should be a free interaction parameter,
rather than a direct signal of the number of active secret streams. If unused
slots were omitted whenever few streams remained, the batch size would expose
the active-stream count. Likewise, if Alice stopped a response immediately
when its payload ended, response length could reveal the fragment boundary or
completion of a stream. Therefore, unused slots are required to look like
ordinary occupied slots and HiTMS allows every occupied response to continue after its real payload is exhausted.

\paragraph{Generality of embedding and extraction}
HiTMS is designed to be \textit{agnostic} both to the LM and to the specific steganographic
coder. 
HiTMS consequently treats the token-level embedder and extractor as black
boxes and adds multi-stream scheduling only through their sequence-level
wrappers. Switching the (LM, coder) pair should not require any
change to the remaining protocol.

\subsection{Multi-Stream Multi-Round Designs}
\label{sec:batch-formulation}

\paragraph{Streams and rounds}
Alice holds $M$ independent secret messages
$\boldsymbol{m}_{i}\in\mathcal{B}^{L_i}$, $i=1,\ldots,M$, which we call
\emph{streams}. A session proceeds over $R$ \emph{rounds}; $R$ is determined by
the time required to exhaust all streams and need not be fixed in advance. In
round $r$, Bob issues a batch of $n_r$ prompts
$\boldsymbol{X}_r=\{\boldsymbol{x}_{r,1},\ldots,
\boldsymbol{x}_{r,n_r}\}$, and Alice returns a corresponding batch of
stegotexts (responses)
$\boldsymbol{Y}_r=\{\boldsymbol{y}_{r,1},\ldots,
\boldsymbol{y}_{r,n_r}\}$. The response in slot $j$ is a token sequence
$\boldsymbol{y}_{r,j}\in\mathcal{V}^{T_{r,j}}$, whose realized length
$T_{r,j}$ is determined by generation up to an EOS token or a prescribed
maximum length. The prompt count $n_r$ may vary across rounds and is not
required to equal the number of streams that remain active.

\paragraph{Stream-to-slot mapping}
Only an active subset of streams is transmitted in any one round. Let
$\mathcal{A}_r\subseteq\{1,\ldots,M\}$ denote the streams that still have
undelivered bits at the start of round $r$, and let
$\mathcal{S}_r=\{1,\ldots,n_r\}$ denote the available response-slot indices.
HiTMS uses a key-derived schedule to select which active streams are served and
to assign the selected streams to distinct slots:
\begin{equation}
\begin{aligned}
    (\mathcal{A}'_r,\,\pi_r)
    \;&=\;
    \mathsf{PRF}_{K}(\textsf{``map''},\,r,\,\mathcal{A}_r),\\
    \mathcal{A}'_r\subseteq\mathcal{A}_r,\qquad
    |\mathcal{A}'_r|
    \;&=\;
    \min(|\mathcal{A}_r|,\,n_r),\qquad
    \pi_r:\mathcal{A}'_r\hookrightarrow\mathcal{S}_r.
\end{aligned}
\end{equation}
Here $\mathcal{A}'_r$ is the served subset, and the injection $\pi_r$ ensures
that no two served streams occupy the same slot. Throughout this paper,
$\mathsf{PRF}_{K}$ denotes a variable-output-length pseudorandom function. Its
output is used either directly as a keystream or as the seed of a
pseudorandom shuffle that derives $(\mathcal{A}'_r,\pi_r)$. All set-valued
arguments are encoded canonically, for example as sorted stream indices, so
that Alice and Bob obtain identical outputs from identical logical states. The
labels $\textsf{``map''}$, $\textsf{``filler''}$, and $\textsf{``xor''}$
provide domain separation among the different uses of the shared key.

The mapping handles both possible relations between the number of active
streams and the number of available slots. When
$|\mathcal{A}_r|>n_r$, exactly $n_r$ streams are selected and streams outside
$\mathcal{A}'_r$ retain their current offsets while waiting for later rounds.
When $|\mathcal{A}_r|<n_r$, every active stream is served, and the surplus
slots in $\mathcal{S}_r\setminus\pi_r(\mathcal{A}'_r)$ carry \emph{decoy}
responses. The bits driving a decoy slot are obtained from
$\boldsymbol{f}_{r,j}=\mathsf{PRF}_{K}(\textsf{``filler''},r,j)$, so the public
batch size need not reveal how many streams are active.

\paragraph{Per-fragment payload}
Consider a served stream $i\in\mathcal{A}'_r$ mapped to slot
$j=\pi_r(i)$. Let
$u_{i,r}=\sum_{r'<r}d_{i,r'}$ be the number of bits of stream $i$ that were
delivered before round $r$, where $d_{i,r'}$ denotes its round-$r'$ payload
and is defined to be zero when the stream is not served in round $r'$. The
residual
$\rho_{i,r}=L_i-u_{i,r}$ is therefore the number of bits still pending when
round $r$ begins. These two quantities partition the message into the already
delivered prefix $\boldsymbol{m}_i[{:}u_{i,r}]$ and the undelivered suffix
$\boldsymbol{m}_i[u_{i,r}{:}L_i]$.

For a served slot, Alice constructs an encrypted header and an encrypted
pending payload. The header encodes the residual payload length \(\rho_{i,r}\) at the beginning of round \(r\), rather than the number of payload bits carried by the current response:
\begin{align}
    \boldsymbol{h}_{i,r}
    \;=&\;
    \mathrm{ENC}_{H}\bigl(\rho_{i,r};\,K,r,j\bigr)
    \;\in\;\mathcal{B}^{\ell_h}, \\
    \boldsymbol{c}_{i,r}
    \;=&\;
    \boldsymbol{m}_{i}\bigl[u_{i,r}{:}L_i\bigr]
    \;\oplus\;
    \mathsf{PRF}_{K}(\textsf{``xor''},\,i)
    \bigl[u_{i,r}{:}L_i\bigr].
\end{align}
The fixed-length header has $\ell_h$ bits; we use $\ell_h=16$, which requires
$L_i<2^{\ell_h}$. The key-derived header encryption
$\mathrm{ENC}_{H}$ is parameterized by the key, round, and slot, and its
decryption allows Bob to recover the advertised residual $\rho_{i,r}$. The
payload is protected by a stream-specific keystream. Slicing this keystream at
$[u_{i,r}{:}L_i]$ ensures that every message position is masked by the
corresponding keystream position even when the stream is fragmented across
several rounds.

The bitstring used to drive the sampler in slot $j$ is
\begin{equation}
\label{eq: The bitstring used to drive the sampler}
\boldsymbol{b}_{r,j}
\;=\;
    \begin{cases}
        \boldsymbol{h}_{i,r}\,\Vert\,\boldsymbol{c}_{i,r},
        & j=\pi_r(i)\ \text{for some}\ i\in\mathcal{A}'_r,\\
        \boldsymbol{f}_{r,j},
        & j\in\mathcal{S}_r\setminus\pi_r(\mathcal{A}'_r).
    \end{cases}
\end{equation}
Equation~\eqref{eq: The bitstring used to drive the sampler} displays the header--payload prefix of an active slot.
Operationally, Alice appends the filler keystream $\boldsymbol{f}_{r,j}$ after
$\boldsymbol{c}_{i,r}$ (as made explicit in Algorithm~\ref{alg:HiTMS-send}), so that the sampler always has sufficient
input bits. Thus, a served slot has the structure
header $\Vert$ payload $\Vert$ filler, whereas a decoy slot contains only
filler. The encrypted active prefix and the decoy keystream are
pseudorandom-looking to a party without $K$, preventing that party from
distinguishing the slot types from their embedded bitstrings.

\paragraph{Embedding and extraction}
For each $j\in\mathcal{S}_r$, Alice runs a sequence-level embedder. All slots
are advanced in batch until each has generated an EOS token or reached the
length cap $T_{\max}$:
\begin{equation}
    \bigl(\boldsymbol{y}_{r,j},\,\beta_{r,j}\bigr)
    \;=\;
    \mathsf{Embed}\bigl(\boldsymbol{x}_{r,j},\,
    \boldsymbol{b}_{r,j},\,K,\,r,\,j\bigr).
\end{equation}
Here $\mathsf{Embed}$ is the autoregressive, sequence-level wrapper of the
token-level coder $\mathsf{Emb}$ in Equation~\eqref{eq: token-level embedder}. It repeatedly evaluates the LM,
invokes $\mathsf{Emb}$, appends the resulting token, and advances the bit
pointer for that slot. The value $\beta_{r,j}$ records the number of body bits
consumed after the $\ell_h$-bit header. It may include both payload and filler
bits, but it does not include the header itself.

As an EOS token may appear at different positions in different responses,
$\beta_{r,j}$ is known only after generation. The real payload
delivered for stream $i$ in round $r$ is
\begin{equation}
    d_{i,r}
    \;=\;
    \min\bigl(\rho_{i,r},\,\beta_{r,\pi_r(i)}\bigr)
    \;\leq\;\rho_{i,r}.
\end{equation}
If the body capacity is smaller than $\rho_{i,r}$, all consumed body bits are
payload and the remaining suffix is deferred. If the body capacity exceeds
$\rho_{i,r}$, the fragment completes the stream and the excess capacity is
filled from $\boldsymbol{f}_{r,j}$.
Across a completed session, the fragments satisfy $\sum_r d_{i,r}=L_i$ for every
stream $i$. Importantly, generation is not terminated at the payload boundary,
so the realized response length does not directly disclose when a stream
finishes.

After receiving $\boldsymbol{Y}_r$, Bob reconstructs
$(\mathcal{A}'_r,\pi_r)$ from his active-set state and the shared key. He then
applies the matching sequence-level extractor to every slot:
\begin{equation}
    \hat{\boldsymbol{b}}_{r,j}
    \;=\;
    \mathsf{Extract}\bigl(\boldsymbol{x}_{r,j},\,
    \boldsymbol{y}_{r,j},\,K,\,r,\,j\bigr),
    \qquad j\in\mathcal{S}_r.
\end{equation}
The wrapper $\mathsf{Extract}$ replays the LM distribution at each observed
token and invokes $\mathsf{Ext}$ from Equation~\eqref{eq: token-level extractor}, concatenating the recovered
substrings in token order. Because the coder is lossless, the resulting
$\hat{\boldsymbol{b}}_{r,j}$ contains exactly the bit prefix consumed by the
embedder.

For a served slot $j=\pi_r(i)$, Bob first decrypts the initial $\ell_h$ bits to
recover $\rho_{i,r}$. He treats the remaining extracted substring as the body
and keeps only its first
$\min(\rho_{i,r},|\hat{\boldsymbol{b}}_{r,j}|-\ell_h)$ bits as encrypted
payload. Applying the corresponding slice of
$\mathsf{PRF}_{K}(\textsf{``xor''},i)$ recovers the next fragment.
Any subsequent body bits are filler and are discarded. Entire decoy slots are
also discarded according to the reconstructed mapping. No explicit field for
$d_{i,r}$ is required: lossless extraction determines how many body bits were
consumed, while the decrypted residual header determines how many of them can
belong to the stream.

The session terminates at
$R^{\star}=\min\{r:\mathcal{A}_{r+1}=\varnothing\}$, namely, after every
residual has reached zero. Conventional single-stream steganography is
recovered as the special case $M=1$ and $n_r=1$ in every round, without the
length header, filler bits, or decoy slots. HiTMS therefore strictly
generalizes the single-stream setting while retaining the original
token-level coder.
\begin{algorithm}[t]\footnotesize
\caption{HiTMS Sender (Alice): Round $r$}
\label{alg:HiTMS-send}
\begin{algorithmic}[1]
\Require Master key $K$; Embedder $\mathsf{Embed}$ with sampling
         hyper-parameters; Header length $\ell_h$; Round index $r$;
         Prompts
         $\boldsymbol{X}_r=\{\boldsymbol{x}_{r,j}\}_{j=1}^{n_r}$;
         Secrets
         $\{\boldsymbol{m}_i\in\mathcal{B}^{L_i}\}_{i=1}^{M}$,
         where $0<L_i<2^{\ell_h}$;
         Offsets $\{u_i\}_{i=1}^{M}$, where $u_i$ is the number of
         bits of $\boldsymbol{m}_i$ already sent;
         Residual lengths
         $\{\rho_i=L_i-u_i\}_{i=1}^{M}$
\Ensure Stegotexts $\boldsymbol{Y}_r$;
        Updated offsets $\{u_i\}$ and residual lengths $\{\rho_i\}$

\State $\mathcal{A}_r\gets\{i:\rho_i>0\}$
\State $(\mathcal{A}'_r,\pi_r)\gets
       \mathsf{PRF}_K(\textsf{``map''},r,\mathcal{A}_r)$

\For{$j=1$ to $n_r$}
    \If{$j=\pi_r(i)$ for some $i\in\mathcal{A}'_r$}
        \State $\boldsymbol{b}_{r,j}\gets
        \mathrm{ENC}_H(\rho_i;K,r,j)
        \,\Vert\,
        \Bigl(
        \boldsymbol{m}_i[u_i{:}L_i]
        \oplus
        \mathsf{PRF}_K(\textsf{``xor''},i)[u_i{:}L_i]
        \Bigr)
        \,\Vert\,
        \mathsf{PRF}_K(\textsf{``filler''},r,j)$
        \Comment{Header $\Vert$ Payload $\Vert$ Filler}
    \Else
        \State $\boldsymbol{b}_{r,j}\gets
        \mathsf{PRF}_K(\textsf{``filler''},r,j)$
        \Comment{Decoy}
    \EndIf
\EndFor

\State $(\{\boldsymbol{y}_{r,j}\},\{\beta_{r,j}\})\gets
       \Call{BatchEmbed}{
       \mathsf{Embed},
       \{\boldsymbol{x}_{r,j}\},
       \{\boldsymbol{b}_{r,j}\},
       K,r}$
\Statex \Comment{$\beta_{r,j}$ is the number of consumed bits after
a complete $\ell_h$-bit header; it is set to $0$ if the header is incomplete}

\For{$i\in\mathcal{A}'_r$ with $j=\pi_r(i)$}
    \State $d\gets\min(\rho_i,\beta_{r,j})$
    \State $u_i\mathrel{+}=d$
    \State $\rho_i\mathrel{-}=d$
\EndFor

\State \Return
       $\boldsymbol{Y}_r=\{\boldsymbol{y}_{r,j}\}_{j=1}^{n_r}$,
       $\{u_i\}_{i=1}^{M}$,
       $\{\rho_i\}_{i=1}^{M}$
\end{algorithmic}
\end{algorithm}

\begin{algorithm}[t]\footnotesize
\caption{HiTMS Receiver (Bob): Round $r$}
\label{alg:HiTMS-recv}
\begin{algorithmic}[1]
\Require Master key $K$; Extractor $\mathsf{Extract}$ with sampling
         hyper-parameters; Header length $\ell_h$; Round index $r$;
         Known stream-index set $\{1,\ldots,M\}$;
         Observed prompts
         $\boldsymbol{X}_r=\{\boldsymbol{x}_{r,j}\}_{j=1}^{n_r}$;
         Stegotexts
         $\boldsymbol{Y}_r=\{\boldsymbol{y}_{r,j}\}_{j=1}^{n_r}$;
         Offsets $\{u_i\}_{i=1}^{M}$, where $u_i$ is the number of
         bits of stream $i$ already recovered;
         Residual lengths $\{\rho_i\}_{i=1}^{M}$,
         where $\rho_i=\bot$ if stream $i$ is unseen;
         Recovered prefixes
         $\{\hat{\boldsymbol{m}}_i\}_{i=1}^{M}$
\Ensure Updated recovered prefixes $\{\hat{\boldsymbol{m}}_i\}$,
        offsets $\{u_i\}$, and residual lengths $\{\rho_i\}$

\State $\mathcal{A}_r\gets
       \{i:\rho_i>0\ \text{or}\ \rho_i=\bot\}$
\State $(\mathcal{A}'_r,\pi_r)\gets
       \mathsf{PRF}_K(\textsf{``map''},r,\mathcal{A}_r)$
       \Comment{Identical to Sender}

\State $\{\hat{\boldsymbol{b}}_{r,j}\}\gets
       \Call{BatchExtract}{
       \mathsf{Extract},
       \{\boldsymbol{x}_{r,j}\},
       \{\boldsymbol{y}_{r,j}\},
       K,r}$

\For{$i\in\mathcal{A}'_r$ with $j=\pi_r(i)$}
    \If{$|\hat{\boldsymbol{b}}_{r,j}|<\ell_h$}
        \State \textbf{continue}
        \Comment{Incomplete header; preserve the current stream state}
    \EndIf

    \State $\hat{\rho}\gets
       \mathrm{ENC}_H^{-1}
       \bigl(
       \hat{\boldsymbol{b}}_{r,j}[{:}\ell_h];
       K,r,j
       \bigr)$
    \State $\boldsymbol{w}\gets
       \hat{\boldsymbol{b}}_{r,j}[\ell_h{:}]$
    \State $t\gets\min(\hat{\rho},|\boldsymbol{w}|)$

    \State $\hat{\boldsymbol{m}}_i\mathrel{\Vert}=
       \boldsymbol{w}[{:}t]
       \oplus
       \mathsf{PRF}_K(\textsf{``xor''},i)
       [u_i{:}u_i{+}t]$

    \State $u_i\mathrel{+}=t$
    \State $\rho_i\gets\hat{\rho}-t$
\EndFor

\State \Return
       $\{\hat{\boldsymbol{m}}_i\}_{i=1}^{M}$,
       $\{u_i\}_{i=1}^{M}$,
       $\{\rho_i\}_{i=1}^{M}$
\end{algorithmic}
\end{algorithm}
\subsection{Per-Round Procedures}

Algorithms~\ref{alg:HiTMS-send} and~\ref{alg:HiTMS-recv} specify one round of
the sender and receiver procedures. A full session initializes the shared
stream index space and repeats these procedures for $r=1,2,\ldots$ until the
active set is empty. Alice initializes $u_i=0$ and $\rho_i=L_i$ for every
stream. Bob initializes $u_i=0$, an empty recovered prefix
$\hat{\boldsymbol{m}}_i$, and $\rho_i=\bot$. The value $\bot$ means that stream
$i$ has not yet appeared in a served slot and that Bob has not yet learned its
residual length. Such an unseen stream is retained in Bob's active set so that
his mapping input agrees with Alice's.

\paragraph{Sender procedure}
At the beginning of round $r$, Alice forms
$\mathcal{A}_r=\{i:\rho_i>0\}$ and derives
$(\mathcal{A}'_r,\pi_r)$ from the shared key. For every served slot, she
serializes the current residual into the encrypted $\ell_h$-bit header, masks
the undelivered message suffix beginning at $u_i$, and appends a filler
keystream. For every surplus slot, she supplies a pure filler keystream. These
$n_r$ bitstrings are passed together with the $n_r$ prompts to
$\textsc{BatchEmbed}$. Internally, this subroutine maintains an independent
coder state and bit pointer for each slot while evaluating the corresponding
LM prefixes in batch. Once all responses end, it returns both the stegotexts
and the realized body-bit counts $\beta_{r,j}$. Alice updates only the streams
in $\mathcal{A}'_r$; waiting streams retain their offsets and residuals.

\paragraph{Receiver procedure}
Bob forms his active set by including every stream whose residual is positive
or still unknown. Because both parties use the same canonical active-set
encoding, round index, and key, Bob derives the same served subset and the same
injection $\pi_r$. He applies $\textsc{BatchExtract}$ to all observed
prompt--response pairs, including the slots that the mapping identifies as
decoys. Extraction itself is uniform across slots; the distinction is made
only when the recovered bitstrings are interpreted. For each served stream,
Bob decrypts the header, sets $\boldsymbol{w}$ to the extracted body, and lets
$t=\min(\hat{\rho},|\boldsymbol{w}|)$. He decrypts exactly the first $t$ body
bits with the stream keystream slice starting at $u_i$, appends the result to
$\hat{\boldsymbol{m}}_i$, advances $u_i$ by $t$, and stores the remaining
residual $\hat{\rho}-t$. All later body bits and all decoy-slot bits are
discarded.

\subsection{Multi-Round Procedures}
\label{sec:multi-round-procedures}

Algorithms~\ref{alg:HiTMS-send-session} and
\ref{alg:HiTMS-recv-session} lift the per-round procedures into complete
interactive sessions. In every round, Bob first supplies a nonempty prompt
batch, Alice invokes Algorithm~\ref{alg:HiTMS-send} once and returns the
resulting stegotext batch, and Bob invokes Algorithm~\ref{alg:HiTMS-recv} once
on the received batch. The next round begins only if the locally reconstructed
active set is nonempty. For compactness, arguments that remain fixed throughout
the session are suppressed when the per-round algorithms are called below.

\begin{algorithm}[t]\footnotesize
\caption{HiTMS Sender (Alice): Multi-Round Session}
\label{alg:HiTMS-send-session}
\begin{algorithmic}[1]
\Require Master key $K$; Embedder $\mathsf{Embed}$ with sampling
         hyper-parameters; Header length $\ell_h$;
         Secrets $\{\boldsymbol{m}_i\in\mathcal{B}^{L_i}\}_{i=1}^{M}$,
         where $0<L_i<2^{\ell_h}$
\Ensure Completion round $R^{\star}$ and terminal sender state

\For{$i=1$ to $M$}
    \State $u_i\gets 0$
    \State $\rho_i\gets L_i$
\EndFor
\State $r\gets 1$

\While{$\exists i\in\{1,\ldots,M\}:\rho_i>0$}
    \State \textbf{receive} a nonempty prompt batch
           $\boldsymbol{X}_r=
           \{\boldsymbol{x}_{r,j}\}_{j=1}^{n_r}$
           from Bob
    \State $(\boldsymbol{Y}_r,\{u_i\},\{\rho_i\})\gets
           \Call{HiTMSSenderRound}{
           r,\boldsymbol{X}_r,\{u_i\},\{\rho_i\}}$
           \Comment{Algorithm~\ref{alg:HiTMS-send}}
    \State \textbf{send} $\boldsymbol{Y}_r$ to Bob
    \State $r\gets r+1$
\EndWhile

\State $R^{\star}\gets r-1$
\State \Return $R^{\star}$,
       $\{u_i\}_{i=1}^{M}$,
       $\{\rho_i\}_{i=1}^{M}$
\end{algorithmic}
\end{algorithm}

\begin{algorithm}[t]\footnotesize
\caption{HiTMS Receiver (Bob): Multi-Round Session}
\label{alg:HiTMS-recv-session}
\begin{algorithmic}[1]
\Require Master key $K$; Extractor $\mathsf{Extract}$ with sampling
         hyper-parameters; Header length $\ell_h$;
         Known stream-index set $\{1,\ldots,M\}$;
         Application-provided prompt-batch source
\Ensure Recovered streams
        $\{\hat{\boldsymbol{m}}_i\}_{i=1}^{M}$
        and completion round $R^{\star}$

\For{$i=1$ to $M$}
    \State $u_i\gets 0$
    \State $\rho_i\gets\bot$
    \State $\hat{\boldsymbol{m}}_i\gets\epsilon$
\EndFor
\State $r\gets 1$

\While{$\exists i\in\{1,\ldots,M\}:
       \rho_i>0\ \text{or}\ \rho_i=\bot$}
    \State $\boldsymbol{X}_r\gets
           \Call{ObtainPromptBatch}{r}$
           \Comment{$\boldsymbol{X}_r\neq\varnothing$}
    \State $n_r\gets|\boldsymbol{X}_r|$
    \State \textbf{send} $\boldsymbol{X}_r$ to Alice
    \State \textbf{receive}
           $\boldsymbol{Y}_r=
           \{\boldsymbol{y}_{r,j}\}_{j=1}^{n_r}$
           from Alice
    \State \parbox[t]{0.94\linewidth}{%
           $(\{\hat{\boldsymbol{m}}_i\},
           \{u_i\},\{\rho_i\})\gets$\\[-0.25ex]
           \hspace*{\algorithmicindent}
           $\Call{HiTMSReceiverRound}{
           r,\boldsymbol{X}_r,\boldsymbol{Y}_r,
           \{u_i\},\{\rho_i\},
           \{\hat{\boldsymbol{m}}_i\}}$}
    \Statex \Comment{Algorithm~\ref{alg:HiTMS-recv}}
    \State $r\gets r+1$
\EndWhile

\State $R^{\star}\gets r-1$
\State \Return
       $\{\hat{\boldsymbol{m}}_i\}_{i=1}^{M}$,
       $R^{\star}$
\end{algorithmic}
\end{algorithm}

\paragraph{Round-by-round synchronization}
The two session loops remain synchronized by induction over rounds. Initially,
Alice has $u_i=0$ and $\rho_i=L_i>0$, whereas Bob has $u_i=0$ and
$\rho_i=\bot$ for every $i$. Thus, both parties include every stream index in
$\mathcal{A}_1$ and derive the same $(\mathcal{A}'_1,\pi_1)$. More generally,
suppose that their active sets and offsets agree at the beginning of round
$r$. They then provide identical inputs $(K,r,\mathcal{A}_r)$ to the mapping
function and obtain the same served subset and slot assignment.

For a served slot with a complete header, losslessness of the underlying coder
gives Bob exactly the header and body prefix consumed by Alice. Hence,
$\hat{\rho}=\rho_i$ and $|\boldsymbol{w}|=\beta_{r,j}$, which gives
\[
t=\min(\hat{\rho},|\boldsymbol{w}|)
 =\min(\rho_i,\beta_{r,j})
 =d.
\]
Both parties apply the same keystream slice at the same offset and advance
that offset by the same amount. If the response contains only a partial
header, both algorithms perform a zero-payload update: Alice has
$\beta_{r,j}=d=0$, while Bob preserves the current stream state. Streams not
selected in the current round are likewise unchanged at both parties.

Consequently, after every completed round,
$\hat{\boldsymbol{m}}_i=\boldsymbol{m}_i[{:}u_i]$; the residuals agree for
every previously seen stream; and an unseen state $\rho_i=\bot$ at Bob
corresponds to a still-undelivered stream at Alice. Their next-round active
sets are therefore identical, which closes the induction.

\paragraph{Batching, framing, and termination}
Each round uses one $\textsc{BatchEmbed}$ call and one matching
$\textsc{BatchExtract}$ call, irrespective of how many of the $n_r$ slots
contain active streams. Header repetition makes every served fragment locally
interpretable, filler permits a response to continue after its payload ends,
and decoys separate the public batch size from the number of active streams.
These mechanisms introduce framing overhead, but they are external to
$(\mathsf{Emb},\mathsf{Ext})$ and therefore preserve coder generality.

After Alice sends the final batch $\boldsymbol{Y}_{R^{\star}}$, her update
makes every residual zero. Lossless extraction makes Bob perform the same
payload updates after receiving that batch, so his loop condition becomes
false in the same round. Bob therefore does not issue
$\boldsymbol{X}_{R^{\star}+1}$, and Alice does not wait for another prompt
batch. No separate termination message or per-fragment payload-length field
is needed.

Finite termination assumes that every unfinished stream is eventually
selected and that selected streams obtain positive body capacity after a
complete header sufficiently often. Under this standard progress condition,
$\sum_r d_{i,r}=L_i$ for every $i$, both algorithms return the same
$R^{\star}$, and Bob returns
$\hat{\boldsymbol{m}}_i=\boldsymbol{m}_i$ for all streams.
\section{Security Analysis of HiTMS}
\label{sec:security-analysis}

We analyze whether the multi-stream framing and scheduling introduced by
HiTMS preserve the security of the underlying single-stream steganographic
coder. The analysis separates three related properties. First,
\emph{steganographic imperceptibility} requires that the responses generated
by HiTMS be indistinguishable from ordinary LM responses. Second,
\emph{stream-pattern privacy} requires that an observer cannot determine
which slots are active, which slots are decoys, or where the payload of an
active slot ends. Third, \emph{message confidentiality} requires that the
secret bits remain hidden even if the bitstrings recovered from the
stegotexts are exposed. Recoverability is addressed separately by the
round-by-round invariant in Section~\ref{sec:multi-round-procedures}. After
introducing a common model and assumptions, the following three main
subsections analyze these properties in the same order.

\subsection{Security Model and Assumptions}

\noindent\textbf{Adversarial setting.}
We consider a probabilistic polynomial-time passive warden $\mathcal{W}$. The
warden knows the LM, tokenizer, steganographic coder, sampling parameters,
header length, and HiTMS protocol, and observes all prompts, batch sizes, and
responses, including their token lengths. It does not know $K$ and does not
modify, delete, insert, or reorder messages. The prompt batch in a later round
may depend on the text observed in earlier rounds, so the model also
covers adaptive interaction.

Let
\begin{equation}
    \Gamma_R
    =
    \bigl(R,\{n_r,\boldsymbol{X}_r\}_{r=1}^{R}\bigr)
\end{equation}
denote the public interaction schedule through round $R$. We compare two
experiments conditioned on the same $\Gamma_R$. In the \emph{HiTMS
experiment}, Alice generates every $\boldsymbol{Y}_r$ using the proposed
protocol. In the \emph{cover experiment}, every response is generated by
ordinary LM sampling under the same prompt, sampling parameters, EOS rule,
and length cap $T_{\max}$. 
For a fixed public schedule $\Gamma_R$, define the complete response
text as
\begin{equation}
    \boldsymbol{Y}_{1:R}
    =
    (\boldsymbol{Y}_1,\ldots,\boldsymbol{Y}_R).
\end{equation}
We denote its distributions in the stegotext and covertext experiments by
\begin{align}
    P_{\mathrm{S}}^{\Gamma_R}
    (\boldsymbol{y}_{1:R})
    &=
    \Pr\bigl[
        \boldsymbol{Y}_{1:R}=\boldsymbol{y}_{1:R}
        \mid \Gamma_R,\text{\rm stegotexts}
    \bigr],\\
    P_{\mathrm{C}}^{\Gamma_R}
    (\boldsymbol{y}_{1:R})
    &=
    \Pr\bigl[
        \boldsymbol{Y}_{1:R}=\boldsymbol{y}_{1:R}
        \mid \Gamma_R,\text{\rm covertexts}
    \bigr].
\end{align}
The warden $\mathcal{W}$ is a binary distinguisher that outputs a bit in
$\{0,1\}$, where output $1$ denotes its decision that the observed text
was steganographic, and output $0$ denotes cover.
The warden's distinguishing advantage is
\begin{equation}
\begin{aligned}
    \operatorname{Adv}^{\mathrm{steg}}_{\mathcal{W}}
    =
    \bigl|
    &\Pr_{\boldsymbol{Y}_{1:R}\leftarrow
          P_{\mathrm{S}}^{\Gamma_R}}
          \bigl[
          \mathcal{W}(\Gamma_R,\boldsymbol{Y}_{1:R})=1
          \bigr]\\
    -{}&
    \Pr_{\boldsymbol{Y}_{1:R}\leftarrow
          P_{\mathrm{C}}^{\Gamma_R}}
          \bigl[
          \mathcal{W}(\Gamma_R,\boldsymbol{Y}_{1:R})=1
          \bigr]
    \bigr|.
\end{aligned}
\end{equation}
This definition treats $\Gamma_R$ as public leakage and asks whether the
response contents and lengths reveal any additional evidence of
steganographic communication.

\noindent\textbf{Cryptographic primitives.}
We assume that $\mathsf{PRF}_K$ is a secure variable-output-length
pseudorandom function and that all of its uses are domain separated by their
text labels. For a concrete fixed-length header encryption, the construction
in Section~\ref{sec:batch-formulation} can be instantiated as
\begin{equation}
    \mathrm{ENC}_H(\rho;K,r,j)
    =
    \operatorname{bin}_{\ell_h}(\rho)
    \oplus
    \mathsf{PRF}_K(\textsf{``header''},r,j)[{:}\ell_h],
\end{equation}
where $\operatorname{bin}_{\ell_h}(\rho)$ is the $\ell_h$-bit binary encoding
of $\rho$. Its inverse applies the same XOR mask. The tuples used by the
\textsf{``map''}, \textsf{``header''}, \textsf{``xor''}, and
\textsf{``filler''} domains are encoded unambiguously.

The proof below concerns one HiTMS session under a fresh session key. Within a
session, every pair $(r,j)$ is unique, and the monotonic offset $u_i$ ensures
that no consumed position of the stream-specific \textsf{``xor''} keystream
is reused. If a long-term master key is shared across multiple sessions, a
fresh session key must first be derived from a unique public session
identifier; we return to this point below.

\noindent\textbf{Underlying steganographic coder.}
HiTMS treats $(\mathsf{Embed},\mathsf{Extract})$ as a black box, so its
distributional security must be inherited from that coder. Let
$\boldsymbol{U}_{r,j}$ be an independent uniform bitstream. For every prompt
$\boldsymbol{x}_{r,j}$ and every public history $\mathcal{H}_{r,j}$ preceding
slot $(r,j)$, assume that
\begin{equation}
\begin{aligned}
    D_{\mathrm{KL}}\bigl(
    &P_{\mathsf{Embed}}
      (\boldsymbol{y}_{r,j}\mid
       \mathcal{H}_{r,j},\boldsymbol{x}_{r,j},\boldsymbol{U}_{r,j})
       \,\Vert\\
    &P_{\mathrm{LM}}
      (\boldsymbol{y}_{r,j}\mid
       \mathcal{H}_{r,j},\boldsymbol{x}_{r,j})
    \bigr)
    \leq \delta_{r,j}.
\end{aligned}
\end{equation}
Both distributions include the EOS event and therefore the realized response
length. The value $\delta_{r,j}$ captures only the distortion introduced by
the underlying coder when it is driven by uniform bits. An exactly
distribution-preserving coder has $\delta_{r,j}=0$, whereas an approximate or
finite-precision coder may have $\delta_{r,j}>0$. Batched inference is assumed
to be an implementation-level parallelization of the same per-slot sampling
process and not to introduce shared sampling randomness across slots.

\subsection{Steganographic Imperceptibility}

\noindent\textbf{Lemma 1 (Pseudorandomness of slot bitstrings).}
Conditioned on any preceding text, the consumed prefix of
$\boldsymbol{b}_{r,j}$ in every slot is computationally indistinguishable
from an independent uniform bitstring. This distribution is independent of
whether the slot carries an active stream or a decoy.

\emph{Justification.}
Replace the domain-separated PRF outputs by independent random functions. In
an active slot, the header is masked by fresh uniform bits, the payload is
XORed with a nonreused uniform keystream slice, and the remaining input is an
independent filler stream. Their concatenation is therefore uniform. A decoy
slot contains only an independent filler stream and has the same distribution.
An incomplete header reveals only a prefix of a uniform string and causes no
state update. Replacing the random functions by $\mathsf{PRF}_K$ changes the
warden's view by at most the PRF distinguishing advantage.
\noindent\textbf{Theorem 1 (Security preservation of HiTMS).}
Let $q$ be an upper bound on the number of domain-separated PRF evaluations
made in an $R$-round session, and let
$\epsilon_{\mathrm{PRF}}(q)$ be the aggregate advantage of replacing those
outputs by independent random functions. Under the preceding assumptions,
every probabilistic polynomial-time passive warden satisfies
\begin{equation}
    \operatorname{Adv}^{\mathrm{steg}}_{\mathcal{W}}
    \leq
    \epsilon_{\mathrm{PRF}}(q)
    +
    \sqrt{
        \frac{1}{2}
        \sum_{r=1}^{R}\sum_{j=1}^{n_r}\delta_{r,j}
    }.
\end{equation}

\emph{Proof sketch.}
Begin with the real HiTMS experiment. In the first hybrid, replace all outputs
in the \textsf{``map''}, \textsf{``header''}, \textsf{``xor''}, and
\textsf{``filler''} domains by independent random-function outputs. A warden
that distinguishes this hybrid from the real experiment with advantage
greater than $\epsilon_{\mathrm{PRF}}(q)$ yields a PRF distinguisher. By
Lemma~1, every slot in the resulting hybrid is driven by an independent
uniform bitstream, regardless of the internal stream pattern.

Next, replace the response of each slot, in text order, by an ordinary
LM response generated under the same prompt and stopping rule. The chain rule
for KL divergence applies to these conditional, potentially adaptive
distributions and gives
\begin{equation}
    D_{\mathrm{KL}}
    \bigl(P_{\mathrm{S}}^{\Gamma_R,\mathrm{ideal}}
    \,\Vert\,
    P_{\mathrm{C}}^{\Gamma_R}\bigr)
    \leq
    \sum_{r=1}^{R}\sum_{j=1}^{n_r}\delta_{r,j}.
\end{equation}
Pinsker's inequality bounds the statistical distance between the ideal hybrid
and the cover experiment by the square-root term in Theorem~1. Adding the PRF
hybrid gap proves the stated bound.

\noindent\textbf{Corollary 1 (Exact distribution preservation).}
If the underlying coder is exactly distribution preserving for uniform driver
bits, so that $\delta_{r,j}=0$ for every slot, then HiTMS has zero KL
divergence from the cover experiment in the random-function model. With a
real secure PRF, the two experiments are computationally indistinguishable
with advantage at most $\epsilon_{\mathrm{PRF}}(q)$. Thus, a statistical
claim applies to the ideal random-function setting, while the corresponding
real-world claim is computational.

\subsection{Stream-Pattern Privacy}

\noindent\textbf{Lemma 2 (Hiding of the stream pattern).}
Fix a public schedule $\Gamma_R$. In the random-function hybrid, the joint
distribution of all slot bitstrings is independent of
$\{\mathcal{A}_r,\mathcal{A}'_r,\pi_r\}_{r=1}^{R}$, the message contents, and
the payload boundaries.

\emph{Justification.}
By Lemma~1, every slot is driven by an independent uniform bitstream,
regardless of its internal role. Permuting active streams among slots,
replacing an active slot with a decoy, changing a covertext message, or moving
the payload--filler boundary therefore leaves the joint driver distribution
unchanged. Because the sequence-level coder receives only the prompt, its
local state, and this driver stream, none of these hidden choices affects the
distribution passed to the LM sampler. In particular, filler ensures that
the bitstream does not end at the payload boundary, so the EOS time and
response length are governed by the same coder-induced distribution for both
active and decoy slots. Replacing the random functions by the real PRF makes
the two internal stream patterns computationally indistinguishable, up to the
corresponding PRF advantage.

\subsection{Message Confidentiality}
The same hybrid argument also protects the message contents. Even if the
warden is conservatively granted the encrypted body bits extracted from each
response, those bits are the XOR of the secret-message fragment and a fresh,
nonreused pseudorandom keystream slice. Therefore, for any two message vectors
consistent with the same public leakage, the resulting encrypted payloads are
computationally indistinguishable. The encrypted residual headers similarly
hide the per-fragment residual lengths. This confidentiality property is
separate from covertness: payload encryption protects what is communicated,
whereas Theorem~1 protects whether steganographic communication is taking
place.


\begin{table*}[t]
\centering
\caption{Comparison between basic 1-stream steganography and our 8-stream HiTMS and 1-stream HiTMS.
Each cell reports mean\,$\pm$\,standard deviation over samples.
Recoverability is omitted, as all stegotexts are correctly extracted (100\%).
Imperceptibility (naturalness and coherence) is rated by an LLM-as-a-judge on a 1--5 scale.}
\label{tab:main_results}
\setlength{\tabcolsep}{2pt}
\renewcommand{\arraystretch}{0.9}
\resizebox{0.9\textwidth}{!}{%
\begin{tabular}{llll cccc cc}
\toprule
\multirow{2}{*}{Dataset} & \multirow{2}{*}{Model} & \multirow{2}{*}{Coder} & \multirow{2}{*}{Method}
& \multicolumn{2}{c}{Capacity} & \multicolumn{2}{c}{Throughput (bits/s)} & \multicolumn{2}{c}{Imperceptibility} \\
\cmidrule(lr){5-6} \cmidrule(lr){7-8} \cmidrule(lr){9-10}
& & & & Bits/Token $\uparrow$ & Payload Util.\ (\%) $\uparrow$
& Embedding $\uparrow$ & Extraction $\uparrow$ & Naturalness\ $\uparrow$ & Coherence\ $\uparrow$ \\
\midrule
\multirow{12}{*}{\makecell[l]{databricks-\\dolly-15k}}
& \multirow{6}{*}{\makecell[l]{Llama-3.2-\\3b-Instruct}}
 & \multirow{3}{*}{AC}
   & $\text{Basic}_{\text{(1-stream)}}$ & 2.35\,\sd{0.66} & 100.0\,\sd{0.0} & 104.2\,\sd{29.2} & 114.2\,\sd{31.9} & 3.17\,\sd{0.89} & 2.90\,\sd{1.16} \\
& & & $\text{HiTMS}_{\text{(8-stream)}}$ & 2.40\,\sd{0.18} & 75.3\,\sd{4.4} & \textbf{442.9}\,\sd{80.5}~(\textbf{4.3}$\times$) & \textbf{495.4}\,\sd{90.7}~(\textbf{4.3}$\times$) & \textbf{3.20}\,\sd{0.90} & \textbf{3.03}\,\sd{1.12} \\
& & & \textcolor{gray!80}{\textit{$\text{HiTMS}_{\text{(1-stream)}}$}} & \textcolor{gray!80}{\textit{2.57\,\sd{0.80}}} & \textcolor{gray!80}{\textit{76.3\,\sd{12.6}}} & \textcolor{gray!80}{\textit{91.9\,\sd{30.0}}} & \textcolor{gray!80}{\textit{101.5\,\sd{33.1}}} & \textcolor{gray!80}{\textit{3.19\,\sd{0.87}}} & \textcolor{gray!80}{\textit{3.15\,\sd{1.10}}} \\
\cmidrule(lr){3-10}
& & \multirow{3}{*}{Discop}
   & $\text{Basic}_{\text{(1-stream)}}$ & 2.61\,\sd{1.03} & 100.0\,\sd{0.0} & 105.2\,\sd{29.0} & 109.4\,\sd{29.7} & 2.79\,\sd{0.97} & 2.70\,\sd{1.16} \\
& & & $\text{HiTMS}_{\text{(8-stream)}}$ & 2.75\,\sd{0.32} & 69.6\,\sd{6.1} & \textbf{294.2}\,\sd{41.2}~(\textbf{2.8}$\times$) & \textbf{304.3}\,\sd{42.4}~(\textbf{2.8}$\times$) & \textbf{2.93}\,\sd{1.05} & \textbf{2.93}\,\sd{1.16} \\
& & & \textcolor{gray!80}{\textit{$\text{HiTMS}_{\text{(1-stream)}}$}} & \textcolor{gray!80}{\textit{3.04\,\sd{1.42}}} & \textcolor{gray!80}{\textit{73.1\,\sd{15.2}}} & \textcolor{gray!80}{\textit{93.5\,\sd{27.8}}} & \textcolor{gray!80}{\textit{97.9\,\sd{28.8}}} & \textcolor{gray!80}{\textit{2.78\,\sd{1.05}}} & \textcolor{gray!80}{\textit{2.81\,\sd{1.20}}} \\
\cmidrule(lr){2-10}
& \multirow{6}{*}{\makecell[l]{Gemma-3-\\4b-it}}
 & \multirow{3}{*}{AC}
   & $\text{Basic}_{\text{(1-stream)}}$ & 0.72\,\sd{0.06} & 100.0\,\sd{0.0} & 18.3\,\sd{1.6} & 20.2\,\sd{1.8} & 4.43\,\sd{0.65} & 4.03\,\sd{1.12} \\
& & & $\text{HiTMS}_{\text{(8-stream)}}$ & 0.72\,\sd{0.02} & 83.6\,\sd{1.3} & \textbf{58.6}\,\sd{6.1}~(\textbf{3.2}$\times$) & \textbf{64.5}\,\sd{6.8}~(\textbf{3.2}$\times$) & \textbf{4.55}\,\sd{0.56} & \textbf{4.14}\,\sd{1.10} \\
& & & \textcolor{gray!80}{\textit{$\text{HiTMS}_{\text{(1-stream)}}$}} & \textcolor{gray!80}{\textit{0.73\,\sd{0.06}}} & \textcolor{gray!80}{\textit{83.9\,\sd{3.4}}} & \textcolor{gray!80}{\textit{15.5\,\sd{1.5}}} & \textcolor{gray!80}{\textit{17.1\,\sd{1.6}}} & \textcolor{gray!80}{\textit{4.55\,\sd{0.54}}} & \textcolor{gray!80}{\textit{4.30\,\sd{1.00}}} \\
\cmidrule(lr){3-10}
& & \multirow{3}{*}{Discop}
   & $\text{Basic}_{\text{(1-stream)}}$ & 0.56\,\sd{0.05} & 100.0\,\sd{0.0} & 14.5\,\sd{1.3} & 15.2\,\sd{1.4} & 4.46\,\sd{0.63} & 4.06\,\sd{1.12} \\
& & & $\text{HiTMS}_{\text{(8-stream)}}$ & 0.57\,\sd{0.01} & 82.0\,\sd{1.0} & \textbf{48.5}\,\sd{4.4}~(\textbf{3.3}$\times$) & \textbf{50.8}\,\sd{4.6}~(\textbf{3.3}$\times$) & \textbf{4.57}\,\sd{0.56} & \textbf{4.26}\,\sd{0.99} \\
& & & \textcolor{gray!80}{\textit{$\text{HiTMS}_{\text{(1-stream)}}$}} & \textcolor{gray!80}{\textit{0.56\,\sd{0.04}}} & \textcolor{gray!80}{\textit{81.9\,\sd{2.8}}} & \textcolor{gray!80}{\textit{12.1\,\sd{1.1}}} & \textcolor{gray!80}{\textit{12.7\,\sd{1.2}}} & \textcolor{gray!80}{\textit{4.55\,\sd{0.56}}} & \textcolor{gray!80}{\textit{4.28\,\sd{1.04}}} \\
\midrule
\multirow{12}{*}{no\_robots}
& \multirow{6}{*}{\makecell[l]{Llama-3.2-\\3b-Instruct}}
 & \multirow{3}{*}{AC}
   & $\text{Basic}_{\text{(1-stream)}}$ & 2.23\,\sd{0.66} & 100.0\,\sd{0.0} & 97.4\,\sd{29.0} & 106.7\,\sd{31.8} & 3.24\,\sd{0.87} & 3.14\,\sd{1.16} \\
& & & $\text{HiTMS}_{\text{(8-stream)}}$ & 2.27\,\sd{0.19} & 75.5\,\sd{4.3} & \textbf{292.3}\,\sd{55.2}~(\textbf{3.0}$\times$) & \textbf{322.0}\,\sd{62.4}~(\textbf{3.0}$\times$) & \textbf{3.45}\,\sd{0.91} & \textbf{3.50}\,\sd{1.08} \\
& & & \textcolor{gray!80}{\textit{$\text{HiTMS}_{\text{(1-stream)}}$}} & \textcolor{gray!80}{\textit{2.41\,\sd{0.80}}} & \textcolor{gray!80}{\textit{77.3\,\sd{11.8}}} & \textcolor{gray!80}{\textit{89.5\,\sd{28.9}}} & \textcolor{gray!80}{\textit{98.9\,\sd{31.9}}} & \textcolor{gray!80}{\textit{3.42\,\sd{0.92}}} & \textcolor{gray!80}{\textit{3.49\,\sd{1.10}}} \\
\cmidrule(lr){3-10}
& & \multirow{3}{*}{Discop}
   & $\text{Basic}_{\text{(1-stream)}}$ & 2.36\,\sd{0.92} & 100.0\,\sd{0.0} & 105.8\,\sd{31.8} & 110.7\,\sd{32.9} & 3.08\,\sd{1.03} & 2.98\,\sd{1.21} \\
& & & $\text{HiTMS}_{\text{(8-stream)}}$ & 2.49\,\sd{0.31} & 70.7\,\sd{6.0} & \textbf{270.6}\,\sd{40.1}~(\textbf{2.6}$\times$) & \textbf{280.4}\,\sd{41.4}~(\textbf{2.5}$\times$) & \textbf{3.14}\,\sd{1.08} & \textbf{3.24}\,\sd{1.21} \\
& & & \textcolor{gray!80}{\textit{$\text{HiTMS}_{\text{(1-stream)}}$}} & \textcolor{gray!80}{\textit{2.65\,\sd{1.24}}} & \textcolor{gray!80}{\textit{73.1\,\sd{14.9}}} & \textcolor{gray!80}{\textit{86.4\,\sd{25.0}}} & \textcolor{gray!80}{\textit{90.5\,\sd{25.9}}} & \textcolor{gray!80}{\textit{3.15\,\sd{1.04}}} & \textcolor{gray!80}{\textit{3.19\,\sd{1.14}}} \\
\cmidrule(lr){2-10}
& \multirow{6}{*}{\makecell[l]{Gemma-3-\\4b-it}}
 & \multirow{3}{*}{AC}
   & $\text{Basic}_{\text{(1-stream)}}$ & 0.64\,\sd{0.07} & 100.0\,\sd{0.0} & 14.1\,\sd{1.5} & 15.4\,\sd{1.6} & 4.38\,\sd{0.75} & 4.08\,\sd{1.04} \\
& & & $\text{HiTMS}_{\text{(8-stream)}}$ & 0.65\,\sd{0.02} & 83.1\,\sd{1.3} & \textbf{57.0}\,\sd{15.7}~(\textbf{4.0}$\times$) & \textbf{63.2}\,\sd{18.5}~(\textbf{4.1}$\times$) & \textbf{4.47}\,\sd{0.70} & \textbf{4.17}\,\sd{1.01} \\
& & & \textcolor{gray!80}{\textit{$\text{HiTMS}_{\text{(1-stream)}}$}} & \textcolor{gray!80}{\textit{0.66\,\sd{0.06}}} & \textcolor{gray!80}{\textit{83.2\,\sd{3.7}}} & \textcolor{gray!80}{\textit{13.8\,\sd{1.5}}} & \textcolor{gray!80}{\textit{15.3\,\sd{1.7}}} & \textcolor{gray!80}{\textit{4.43\,\sd{0.68}}} & \textcolor{gray!80}{\textit{4.15\,\sd{1.02}}} \\
\cmidrule(lr){3-10}
& & \multirow{3}{*}{Discop}
   & $\text{Basic}_{\text{(1-stream)}}$ & 0.50\,\sd{0.05} & 100.0\,\sd{0.0} & 11.6\,\sd{1.3} & 12.1\,\sd{1.3} & 4.35\,\sd{0.78} & 4.01\,\sd{1.12} \\
& & & $\text{HiTMS}_{\text{(8-stream)}}$ & 0.50\,\sd{0.02} & 81.2\,\sd{1.1} & \textbf{39.5}\,\sd{6.5}~(\textbf{3.4}$\times$) & \textbf{41.2}\,\sd{7.0}~(\textbf{3.4}$\times$) & \textbf{4.51}\,\sd{0.65} & \textbf{4.17}\,\sd{0.98} \\
& & & \textcolor{gray!80}{\textit{$\text{HiTMS}_{\text{(1-stream)}}$}} & \textcolor{gray!80}{\textit{0.50\,\sd{0.05}}} & \textcolor{gray!80}{\textit{81.2\,\sd{3.0}}} & \textcolor{gray!80}{\textit{12.4\,\sd{2.7}}} & \textcolor{gray!80}{\textit{13.2\,\sd{3.0}}} & \textcolor{gray!80}{\textit{4.43\,\sd{0.70}}} & \textcolor{gray!80}{\textit{4.12\,\sd{1.00}}} \\
\midrule
\multicolumn{4}{l}{\emph{Average of $\text{Basic}_{\text{(1-stream)}}$}}
   & 1.50 & 100.0 & 58.9 & 63.0 & 3.74 & 3.49 \\
\multicolumn{4}{l}{\emph{Average of $\text{HiTMS}_{\text{(8-stream)}}$}}
   & 1.54 & 77.6 & \textbf{188.0}~(\textbf{3.2}$\times$) & \textbf{202.7}~(\textbf{3.2}$\times$) & \textbf{3.85} & \textbf{3.68} \\
\bottomrule
\end{tabular}}
\end{table*}

\section{Experiments}
\label{sec: experiments}
\subsection{Setup}

To validate the generality of our HiTMS framework, we implemented it using two language models: Llama-3.2-3b-Instruct ~\cite{llama3modelcard} and Gemma-3-4b-it~\cite{gemma_2025}. For the steganographic coders, we adopted the lightweight arithmetic coding, AC~\cite{ziegler-etal-2019-neural} and a typical provably secure method, Discop~\cite{10179287}.
The questions and prompts are drawn from databricks-dolly-15k~\cite{DatabricksBlog2023DollyV2} and no\_robots~\cite{no_robots}. From each dataset, we retained the open-ended, creative instructions, as they elicit sufficiently high-entropy responses for embedding.

At each step the target distribution is the model's temperature-scaled
($\tau = 1.0$) distribution over the full vocabulary (top-$p$, $p=1.0$); the maximum
response length is $T_{\text{max}}=256$ tokens.
$500$ interaction (session) samples were gathered for each experimental group, and all
random seeds (pool shuffling, the per-round batch size $n$, the secret payloads,
and model sampling) were fixed for reproducibility.
The length of the secret message $L_i$ for each stream is $1{,}024$ and each bit is independently and uniformly drawn from $\{0,1\}$.
All experiments were conducted on NVIDIA RTX A6000 (48\,GB) GPUs.

\subsection{Main Metrics}
HiTMS is designed to improve throughput while maintaining recoverability and steganographic imperceptibility. The main metrics are as follows:
\begin{itemize}
    \item \textbf{Capacity:} (1) the number of secret bits embedded per token (bits/token); (2) payload utilization (\%), i.e., the ratio of secret (payload) bits  embedded to total embedded bits  (length header, secret bits, and filler bits).
    \item \textbf{Throughput:} the embedding and extraction speed (bits/s), counting payload bits only.
    \item \textbf{Recoverability:} the proportion (\%) of stegotexts from which the secret message is correctly extracted.\footnote{Tokenization inconsistency and segmentation ambiguity are beyond the scope of this work, as existing countermeasures~\cite{nozaki-murawaki-2022-addressing,yan2023A,10831370,qi2024provably,yan2025addressingtokenizationinconsistencysteganography} are orthogonal to our HiTMS framework.}
    \item \textbf{Imperceptibility:} an LLM-as-a-judge (\textit{gpt-5.4-2026-03-05}) rates each Alice--Bob interaction on a 1 (lowest)--5 (highest) scale along two imperceptibility-related dimensions: (1) answer naturalness and (2) answer coherence (including logical coherence and topical consistency). LLM-as-a-judge metrics~\cite{gu2025surveyllmasajudge} can align more closely with human perception than traditional automatic metrics such as perplexity~\cite{jelinek1977perplexity} or BERTScore~\cite{Zhang*2020BERTScore:}.
\end{itemize}

\subsection{Main Results}
In this section, HiTMS was evaluated with $M = 8$ streams, the prompt number (issued by Bob) $n_r\sim\text{Uniform}\{1,...,M\}$ for each round $r$.
Experiments were also conducted on 1-stream HiTMS for reference and ablation.
To compare our HiTMS against the basic 1-stream steganography (baseline), we additionally conducted 1-stream basic but multi-round linguistic steganography. The two share an identical basic setup, except that basic 1-stream baseline requires no frame design and thus contains no header or filler bits, giving it a constant payload utilization of 100\%.
Table~\ref{tab:main_results} reports the results of capacity, throughput and imperceptibility. The recoverability is not reported separately, as all stegotexts are correctly extracted (recoverability $= 100\%$). From this table, the following observations stand out:

\paragraph{Multi-stream HiTMS delivers substantially higher throughput} By embedding and extracting all streams of a round in a single batched LM call, HiTMS obtains up to $4.3\times$ faster embedding and extraction speeds than the single-stream baseline across all datasets, models, and coders. Since $n_r\sim\text{Uniform}\{1,...,M\}$, a single batched LM call serves $E(n_r)=\frac{M+1}{2}=4.5$ streams on average. However, the observed speedup stays below $4.5\times$, because of the frame overhead.

\paragraph{The frame overhead is modest} HiTMS generally matches the bits per token of the basic 1-stream steganography. The price of the self-describing frame is a payload utilization of roughly $70\%$--$85\%$, reflecting the length headers and filler bits required for recoverability and pattern indistinguishability.

\paragraph{Imperceptibility is not only preserved but even slightly improved} The LLM-as-a-judge naturalness and coherence scores of HiTMS are consistently above those of the single-stream baseline (e.g., naturalness $3.45$ vs.\ $3.24$ for AC on no\_robots with Llama). We attribute this gain to our filler-equipped frame design: the filler bits keep a response going after its payload ends to avoid cutting a sentence off mid-way.

\begin{table}[t]
\centering
\caption{Steganalysis dataset for the universal detector. Covertexts are payload-free
text generated from the same prompts and model. Each of the $16$ test cells
(model$\times$dataset$\times$coder$\times$method) is balanced to $500$ responses.}
\label{tab:steg_data}
\setlength{\tabcolsep}{8pt}
\begin{tabular}{lrrr}
\toprule
Split & Covertexts & Stegotexts & Total \\
\midrule
Train (mixed)      & 11{,}840 & 11{,}840 & 23{,}680 \\
Validation (mixed) &  2{,}896 &  2{,}896 &  5{,}792 \\
Test ($16\times500$) &  4{,}000 &  4{,}000 &  8{,}000 \\
\bottomrule
\end{tabular}
\end{table}

\begin{table*}[t]
\centering
\caption{Anti-steganalysis performance measured as detector AUROC
(\textbf{$\downarrow$ is more secure}; $0.5$ is random). Five pretrained encoders serve as steganalyzers; the last column averages them. In every
($\text{Basic}_{\text{(1-stream)}}$, $\text{HiTMS}_{\text{(8-stream)}}$) pair the more secure (lower-AUROC) data point is in \textbf{bold}.}
\label{tab:steg_main}
\setlength{\tabcolsep}{8pt}
\renewcommand{\arraystretch}{0.9}
\begin{tabular}{llll ccccc c}
\toprule
Dataset & Model & Coder & Method
& BERT$_{\text{base}}$ & BERT$_{\text{large}}$ & RoBERTa & DeBERTa-v3 & ELECTRA & Avg. \\
\midrule
\multirow{8}{*}{\makecell[l]{databricks-\\dolly-15k}}
& \multirow{4}{*}{\makecell[l]{Llama-3.2-\\3b-Instruct}}
 & \multirow{2}{*}{AC}
   & $\text{Basic}_{\text{(1-stream)}}$ & 0.873 & 0.830 & 0.867 & 0.858 & 0.878 & 0.861 \\
& & & $\text{HiTMS}_{\text{(8-stream)}}$   & \textbf{0.796} & \textbf{0.734} & \textbf{0.794} & \textbf{0.779} & \textbf{0.803} & \textbf{0.781} \\
\cmidrule(lr){3-10}
& & \multirow{2}{*}{Discop}
   & $\text{Basic}_{\text{(1-stream)}}$ & 0.837 & 0.796 & 0.844 & 0.830 & 0.845 & 0.830 \\
& & & $\text{HiTMS}_{\text{(8-stream)}}$   & \textbf{0.739} & \textbf{0.678} & \textbf{0.748} & \textbf{0.753} & \textbf{0.733} & \textbf{0.730} \\
\cmidrule(lr){2-10}
& \multirow{4}{*}{\makecell[l]{Gemma-3-\\4b-it}}
 & \multirow{2}{*}{AC}
   & $\text{Basic}_{\text{(1-stream)}}$ & 0.579 & 0.535 & 0.571 & 0.569 & 0.567 & 0.564 \\
& & & $\text{HiTMS}_{\text{(8-stream)}}$   & \textbf{0.481} & \textbf{0.467} & \textbf{0.491} & \textbf{0.483} & \textbf{0.490} & \textbf{0.482} \\
\cmidrule(lr){3-10}
& & \multirow{2}{*}{Discop}
   & $\text{Basic}_{\text{(1-stream)}}$ & 0.527 & 0.506 & 0.505 & 0.562 & 0.542 & 0.528 \\
& & & $\text{HiTMS}_{\text{(8-stream)}}$   & \textbf{0.505} & \textbf{0.459} & \textbf{0.476} & \textbf{0.516} & \textbf{0.524} & \textbf{0.496} \\
\midrule
\multirow{8}{*}{no\_robots}
& \multirow{4}{*}{\makecell[l]{Llama-3.2-\\3b-Instruct}}
 & \multirow{2}{*}{AC}
   & $\text{Basic}_{\text{(1-stream)}}$ & 0.807 & 0.736 & 0.812 & 0.781 & 0.795 & 0.786 \\
& & & $\text{HiTMS}_{\text{(8-stream)}}$   & \textbf{0.707} & \textbf{0.622} & \textbf{0.678} & \textbf{0.704} & \textbf{0.701} & \textbf{0.682} \\
\cmidrule(lr){3-10}
& & \multirow{2}{*}{Discop}
   & $\text{Basic}_{\text{(1-stream)}}$ & 0.767 & 0.692 & 0.782 & 0.746 & 0.747 & 0.747 \\
& & & $\text{HiTMS}_{\text{(8-stream)}}$   & \textbf{0.678} & \textbf{0.588} & \textbf{0.649} & \textbf{0.645} & \textbf{0.653} & \textbf{0.643} \\
\cmidrule(lr){2-10}
& \multirow{4}{*}{\makecell[l]{Gemma-3-\\4b-it}}
 & \multirow{2}{*}{AC}
   & $\text{Basic}_{\text{(1-stream)}}$ & 0.547 & 0.578 & 0.560 & 0.611 & 0.593 & 0.578 \\
& & & $\text{HiTMS}_{\text{(8-stream)}}$   & \textbf{0.501} & \textbf{0.509} & \textbf{0.461} & \textbf{0.512} & \textbf{0.483} & \textbf{0.493} \\
\cmidrule(lr){3-10}
& & \multirow{2}{*}{Discop}
   & $\text{Basic}_{\text{(1-stream)}}$ & 0.551 & 0.566 & 0.525 & 0.572 & 0.561 & 0.555 \\
& & & $\text{HiTMS}_{\text{(8-stream)}}$   & \textbf{0.487} & \textbf{0.484} & \textbf{0.514} & \textbf{0.497} & \textbf{0.498} & \textbf{0.496} \\
\midrule
\multicolumn{4}{l}{\emph{Average of $\text{Basic}_{\text{(1-stream)}}$}}
   & 0.686 & 0.655 & 0.683 & 0.691 & 0.691 & 0.681 \\
\multicolumn{4}{l}{\emph{Average of $\text{HiTMS}_{\text{(8-stream)}}$}}
   & \textbf{0.612} & \textbf{0.568} & \textbf{0.601} & \textbf{0.611} & \textbf{0.611} & \textbf{0.601} \\
\bottomrule
\end{tabular}
\end{table*}

\begin{figure*}[t]
\centering
\includegraphics[width=\textwidth]{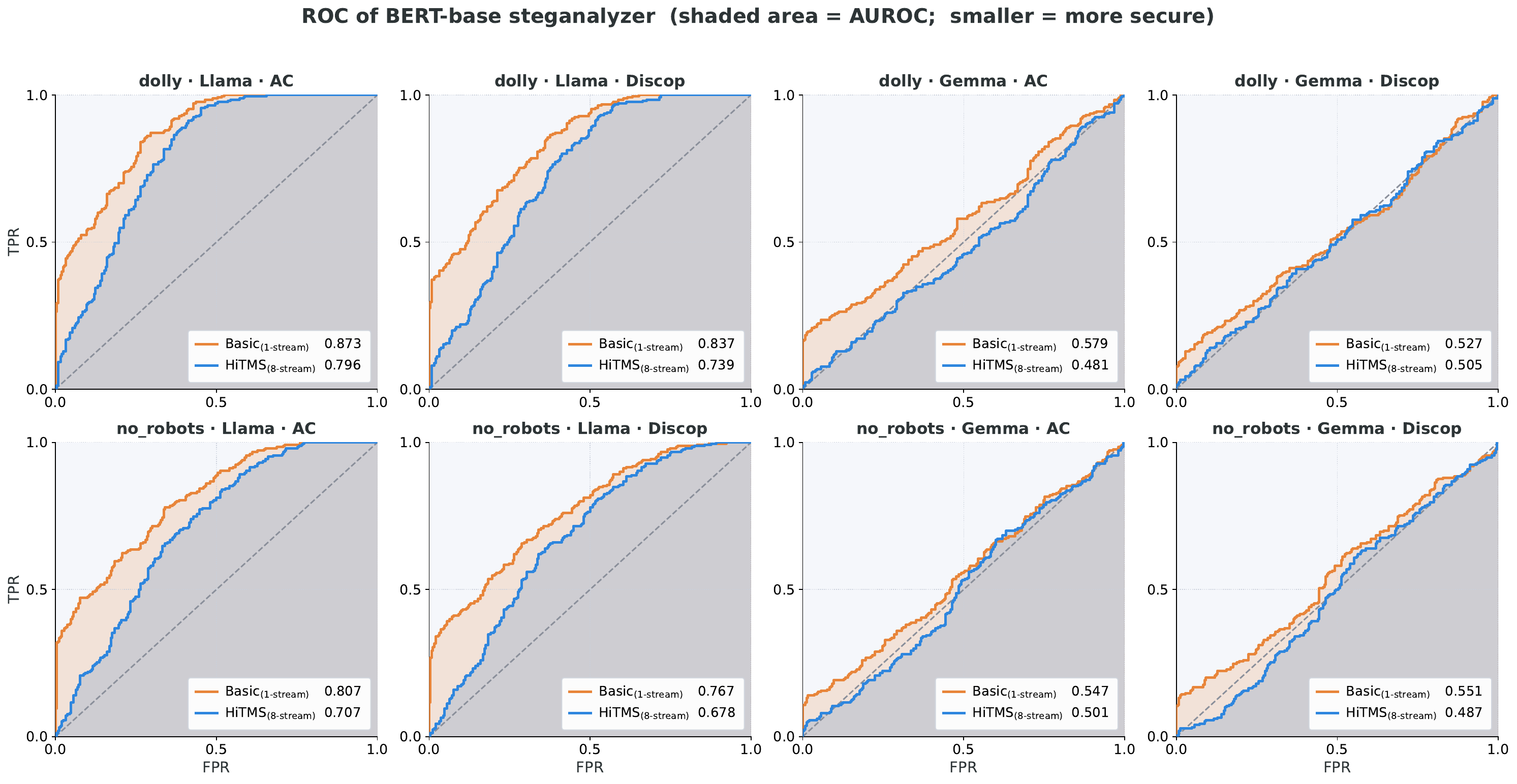}
\caption{Receiver operating characteristic (ROC) curves of the BERT-base
steganalyzer for each of the eight experimental settings
(dataset\,$\times$\,model\,$\times$\,coder), comparing the single-stream
baseline (orange) against our 8-stream HiTMS (blue). Each curve traces the detector's
true positive rate (TPR) against its false positive rate (FPR) as the decision
threshold is swept, where the TPR is the fraction of stegotexts
correctly flagged as steganographic (i.e., recall) and the FPR is the fraction of
payload-free covertexts wrongly flagged as steganographic.}
\label{fig:roc}
\end{figure*}

\subsection{Analyses of Anti-Steganalysis Capability}
\label{sec: Analyses of Anti-Steganalysis Capability}

We assessed security under the standard steganalysis threat model, in which a warden is a binary classifier trained to separate stegotexts from innocent text (covertexts). To isolate the effect of the embedding alone, the cover class was generated with the \emph{same} model, the \emph{same} prompts, and the \emph{same} sampling distribution ($\tau{=}1.0$, full vocabulary, chat-control tokens masked, EOS unmasked) as the steganographic class runs, but \emph{without} embedding any payload, i.e., via plain multinomial sampling. 
For each (model, dataset) pair we generated $4{,}684$ payload-free cover responses that replay the prompts consumed by the corresponding stegotexts run, giving $4$ cover sets ($18{,}736$ responses in total).

We trained a \emph{single universal} detector over a mixture of all configurations and report its per-cell test performance; this matches the realistic case in which the warden does not
know which configuration produced a given text. Within every (model, dataset) block the cover and the steganographic counts are matched and every configuration contributes an equal number of steganographic samples, so model/dataset identity carries no label information.
Samples are split \emph{grouped by question} and each of the $16$ test cells is balanced to exactly $250$ stegotexts with $250$ cover responses (Table~\ref{tab:steg_data} excluding $\text{HiTMS}_{\text{1-stream}}$ rows). 
We used five pretrained encoders as detectors: BERT-base\footnote{\url{https://huggingface.co/google-bert/bert-base-uncased}}~\cite{DBLP:journals/corr/abs-1810-04805}, BERT-large\footnote{\url{https://huggingface.co/google-bert/bert-large-uncased}}~\cite{DBLP:journals/corr/abs-1810-04805}, RoBERTa-base\footnote{\url{https://huggingface.co/FacebookAI/roberta-base}}~\cite{DBLP:journals/corr/abs-1907-11692}, DeBERTa-v3-base\footnote{\url{https://huggingface.co/microsoft/deberta-v3-base}}~\cite{he2021deberta},
and ELECTRA-base\footnote{\url{https://huggingface.co/google/electra-base-discriminator}}~\cite{clark2020electrapretrainingtextencoders}. Each of them is fine-tuned for up to $5$ epochs with AdamW
(learning rate $ = 2{\times}10^{-5}$), batch size $32$,
$\mathrm{max\_len}{=}256$. 
We report AUROC ($\downarrow$ is more secure) with $500$ test samples per cell.

\paragraph{Main detection results}
Table~\ref{tab:steg_main} reports the per-cell AUROC of all five detectors.
Two findings are consistent across every encoder. First, on \mbox{Gemma-3-4b-it} the steganographic text is highly imperceptible: the AUROC on Gemma cells generally lies around the $0.5$ chance level (random guessing), and neither the larger BERT-large nor the stronger DeBERTa-v3 shows a consistent advantage over chance. These results indicate that the detection failure is not one of detector capacity but an absence of exploitable surface artifacts. Second, on \mbox{Llama-3.2-3b-Instruct} the stegotexts are relatively detectable, yet
our multi-stream HiTMS is \emph{consistently harder to detect than the
single-stream} baseline: averaged over the eight Llama cells, HiTMS lowers the
AUROC by $0.08$--$0.11$ relative to single-stream for every encoder, and in all $40$ (setting $\times$ encoder) paired comparisons, HiTMS is harder to detect than the single-stream baseline. Similarly to analyses on imperceptibility metrics, the filler bits keep generating tokens in stegotexts after its payload ends to avoid mid-sentence termination.
Besides, Figure~\ref{fig:roc} reports the ROC curves of the BERT-base steganalyzer for each setting.

\paragraph{Where the residual detectability comes from}
Part of the single-stream detectability is a \emph{length} artifact:
the single-stream protocol truncates its final response the moment the fixed
payload budget is exhausted, so its responses are systematically shorter than
cover, whereas HiTMS spreads the payload over several naturally terminated
responses and leaks almost no length signal. Table~\ref{tab:steg_len} quantifies
this with a logistic-regression detector that uses a \emph{single} feature, the
response length in tokens. For single-stream on Llama this trivial detector
already reaches $0.700$ AUROC (mean length gap $\approx\!40$ tokens), whereas for
HiTMS it collapses toward chance ($0.575$; gap $\approx\!5$ tokens). Thus a
non-trivial share of the single-stream signal is length alone, a cue that HiTMS
structurally removes. We keep this length difference in the data, since under a realistic warden the length distribution is itself an observable and legitimate detection cue.

\begin{table}[t]
\centering
\caption{Trivial length-only detector (logistic regression on response length in tokens), averaged over the four coder/dataset cells of each Model$\times$Method.
Single-stream leaks a length fingerprint that HiTMS removes.}
\label{tab:steg_len}
\setlength{\tabcolsep}{4pt}
\begin{tabular}{llccc}
\toprule
Model & Method & \makecell[l]{Avg. covertext \\ length} & \makecell[l]{Avg. stegotext \\ length} & \makecell[l]{Len-only\\AUROC} \\
\midrule
\multirow{2}{*}{\makecell[l]{Llama-3.2-\\3b-Instruct}}
  & $\text{Basic}_{\text{(1-stream)}}$ & 231 & 192 & 0.700 \\
  & $\text{HiTMS}_{\text{(8-stream)}}$    & 231 & 226 & 0.575 \\
\cmidrule(lr){1-5}
\multirow{2}{*}{\makecell[l]{Gemma-3-\\4b-it}}
  & $\text{Basic}_{\text{(1-stream)}}$ & 221 & 208 & 0.558 \\
  & $\text{HiTMS}_{\text{(8-stream)}}$    & 221 & 222 & 0.498 \\
\bottomrule
\end{tabular}
\end{table}

\paragraph{Coder comparison}
The two coders are ranked consistently: Discop is generally harder to detect than arithmetic coding (AC). 
On Llama-3.2-3b-Instruct, Discop lowers the detector AUROC below AC in \emph{all} $20$ paired comparisons (4 settings $\times$ 5 encoders), by $0.040$ on average
(AC $0.778$ vs.\ Discop $0.738$); the gap is stable across every encoder ($0.032$--$0.049$). 
On Gemma-3-4b-it the same ordering holds in $14/20$ comparisons
(AC $0.530$ vs.\ Discop $0.519$). 
This ordering is consistent with the provable security of Discop, which preserves the token distribution exactly, whereas arithmetic coding perturbs it slightly through integer
quantization of the cumulative distribution which is a residual distortion that a steganalyzer can partially exploit.


\subsection{Analyses of Multi-Stream Scalability}

\begin{figure*}[t]
\centering
\includegraphics[width=0.9\textwidth]{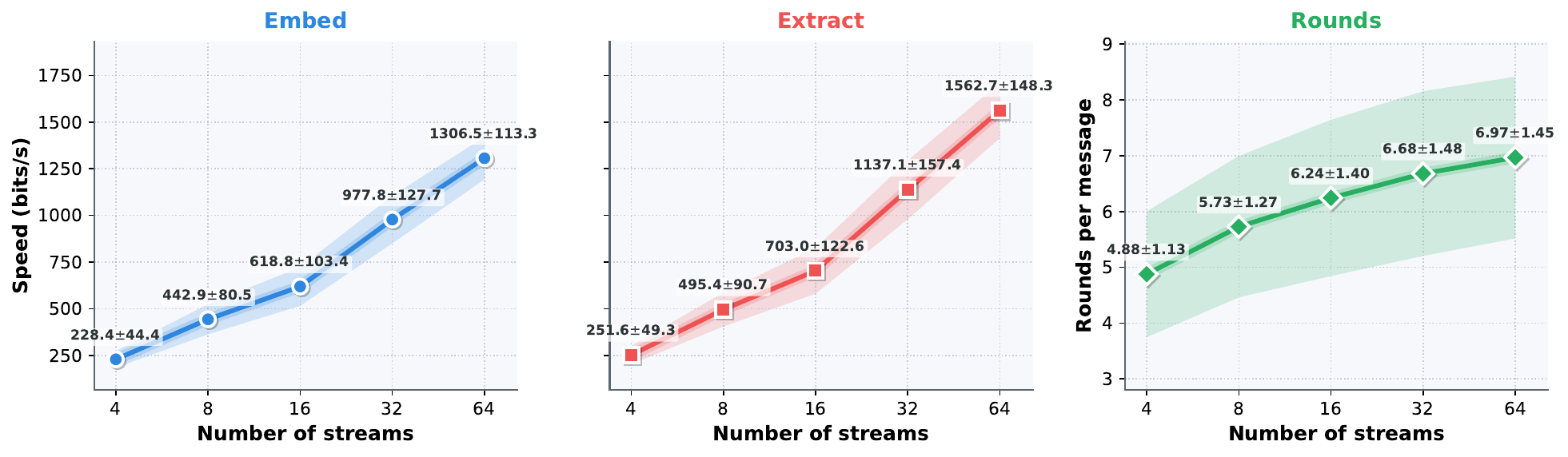}
\caption{Embedding and extraction speed (bits/s) and number of rounds of HiTMS as the number of streams $M$ scales over $\{4,8,16,32,64\}$, on databricks-dolly-15k with Llama-3.2-3b-Instruct and arithmetic coding.}
\label{fig:scaling_speed_arithmetic}
\end{figure*}

\begin{figure*}[t]
\centering
\includegraphics[width=0.9\textwidth]{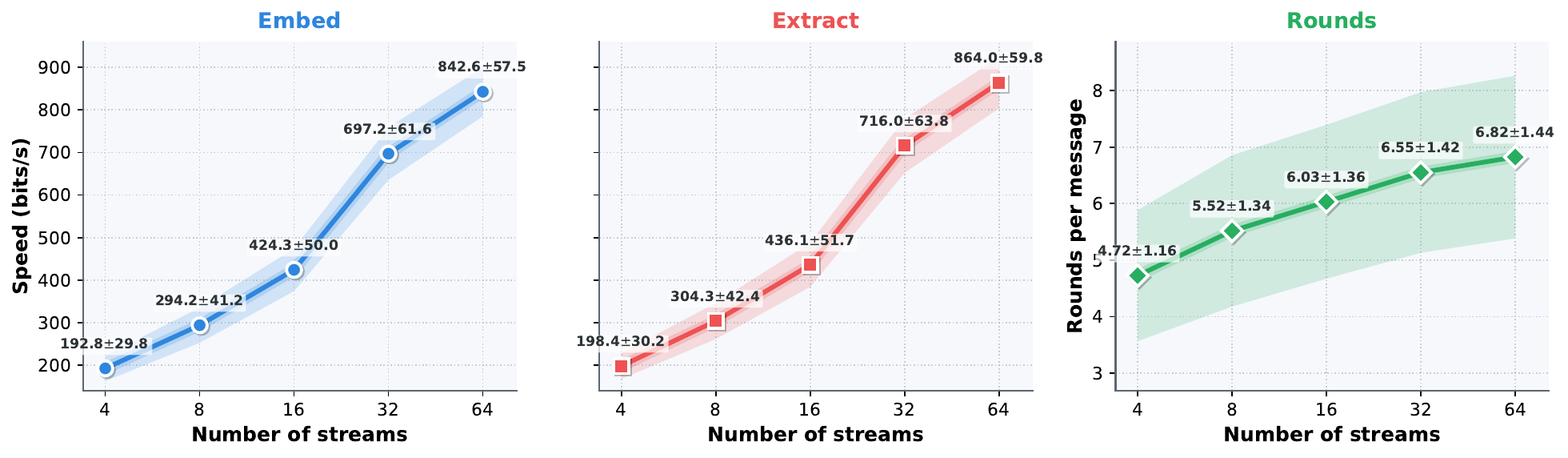}
\caption{Embedding and extraction speed (bits/s) and number of rounds of HiTMS as the number of streams $M$ scales over $\{4,8,16,32,64\}$, on databricks-dolly-15k with Llama-3.2-3b-Instruct and Discop.}
\label{fig:scaling_speed}
\end{figure*}

We further evaluated the multi-stream scalability of HiTMS on databricks-dolly-15k using Llama-3.2-3b-Instruct with both arithmetic coding (AC) and Discop. We varied the number of streams as \(M \in \{4,8,16,32,64\}\) and sampled the number of prompts in each round as \(n_r \sim \operatorname{Uniform}\{1,\ldots,M\}\). Figures~\ref{fig:scaling_speed_arithmetic} and~\ref{fig:scaling_speed} report the embedding speed, extraction speed, and number of rounds for AC and Discop, respectively.

For both coders, the embedding and extraction throughput increases consistently with the number of streams. When \(M\) increases from \(4\) to \(64\), the embedding and extraction speeds of AC increase from \(228.4\) to \(1306.5\) bits/s and from \(251.6\) to \(1562.7\) bits/s, corresponding to \(5.7\times\) and \(6.2\times\) improvements, respectively. Under Discop, the corresponding speeds increase from \(192.8\) to \(842.6\) bits/s and from \(198.4\) to \(864.0\) bits/s, both by approximately \(4.4\times\).
Both coders obtain substantial and consistent throughput improvements as \(M\) increases, confirming the effectiveness of HiTMS for large-scale concurrent transmission.
They demonstrate that the scalability benefit is not specific to a particular coding scheme.

The throughput growth is nevertheless sub-linear relative to the \(16\times\) increase in the number of streams. We attribute this behavior to two factors. First, although batched inference amortizes the cost of LM forward passes, token-level embedding and extraction for every response slot still require coder-specific operations on the CPU. As the batch grows, this per-slot coding overhead becomes increasingly significant and limits the benefit obtained from larger batches. Second, a session terminates only after its last stream has been completely transmitted. Increasing \(M\) raises the probability that at least one stream requires additional rounds, producing a tail-completion effect. Accordingly, the average number of rounds increases from \(4.88\) to \(6.97\) for AC and from \(4.72\) to \(6.82\) for Discop. These additional tail rounds contain fewer valid payload bits after most streams have completed, thereby reducing the average valid throughput per batched call. 

\section{Related Work}
\label{sec: Related Work}

\subsection{Generative Linguistic Steganography}
Generative linguistic steganography embeds secret bits directly into the token-level sampling randomness of an
autoregressive LM, as pioneered by RNN-Stega~\cite{8470163} and VAE-Stega~\cite{9193914}.
Ziegler~\emph{et al.}~\cite{ziegler-etal-2019-neural} introduced arithmetic coding for near-optimal capacity.
ADG~\cite{zhang-etal-2021-provably} groups tokens adaptively to approximate a uniform partition of the distribution. A parallel line pursues \emph{provable security}:
Discop~\cite{10179287} embeds bits via ``distribution copies'' that exactly preserve the model
distribution; Shimmer~\cite{bai2025shimmer} 
collects entropy across steps; SparSamp~\cite{wang2025sparsampefficientprovablysecure} improves efficiency through sparse sampling; and rotation range coding (RRC) has been employed for efficient provably secure embedding~\cite{yan-murawaki-2026-efficient}. 
Despite this diversity of coders and backbones, all of the above auto-regressive schemes are \emph{single-stream}: one secret is conveyed through one
response to one prompt. HiTMS is orthogonal to this line and generalizes the transmission protocol itself to a multi-stream, multi-round setting.
 
\subsection{Recoverability and Robustness}
Reliable extraction is a long-standing concern in LM-based steganography. One family of work addresses extraction failures caused by segmentation ambiguity or tokenization inconsistency, either by disambiguating candidate pools~\cite{nozaki-murawaki-2022-addressing,yan2023A,10831370,qi2024provably} or by enforcing tokenization-consistent generation~\cite{yan2025addressingtokenizationinconsistencysteganography}.
Another family enhances robustness against channel distortions, e.g., the adaptive enhancement framework WinStega~\cite{10888944} and the mechanism of the anchored sliding window~\cite{yan-etal-2026-anchored}.
These methods all safeguard recoverability \emph{within} a single response and are therefore complementary to HiTMS.
 
\subsection{Steganography as a Service and Efficient LLM Inference}
The vision of offering steganography as a cloud service dates back to service-oriented steganography~\cite{5166821} and has been revisited for cloud data security~\cite{8646434,9004347} and, recently, microservice-based video steganography~\cite{11524553}.
However, these efforts largely target image or video carriers and do not address the serving cost of modern LLMs. 
On the systems side, batching is the standard technique for amortizing LLM inference cost, e.g., batch prompting for efficient API usage~\cite{cheng-etal-2023-batch}. To the best of our knowledge, no prior work connects batched auto-regressive LM inference with generative linguistic steganography: existing schemes specify one payload-bearing stream per protocol instance. Although independent instances can be co-batched at the implementation level, they do not provide the cross-round framing, scheduling, and stream-pattern privacy formalized by HiTMS.
By fragmenting secrets across concurrent streams, HiTMS integrates batched inference into a multi-stream, multi-round steganographic protocol and thereby provides the throughput required by concurrent requests for a linguistic version of steganography as a service (LSaaS) in the future.

\section{Conclusion}

We presented HiTMS, a model- and coder-agnostic framework that generalizes conventional single-stream linguistic steganography to multi-stream, multi-round interaction. By distributing secret messages across concurrent response streams, HiTMS makes batched autoregressive LM inference applicable to linguistic steganography and amortizes the cost of model invocation. Its self-describing frames, encrypted headers, key-derived stream-to-slot mapping, filler bits, and decoy responses enable exact recovery while concealing the active-stream pattern. 
Experiments across various datasets, models, and coders show that HiTMS substantially improves throughput while reducing steganalysis detectability. Scalability analyses further confirm sustained throughput gains as concurrency increases.
These results position HiTMS as a promising foundation for a linguistic version of steganography as a service (LSaaS), where multiple concurrent requests need to be handled efficiently without sacrificing recoverability or imperceptibility.

\bibliographystyle{IEEEtran}

\bibliography{references}
\end{document}